\documentclass[onecolumn,preprintnumbers,showkeys,showpacs]{revtex4}
\usepackage{amsmath, amsthm, amscd, amsfonts, amssymb, graphicx, hhline}
\usepackage{dcolumn}
\usepackage{bm}
\usepackage[english]{babel}
\usepackage{footnpag}
\usepackage{nccfoots}
\usepackage{multirow}
\usepackage[]{color}
\usepackage[symbol*]{footmisc}

\begin{document}

\title{Collective excitations in two-band superconductors}
\author{Konstantin V. Grigorishin}

\email{konst.phys@gmail.com} \affiliation{Bogolyubov Institute for Theoretical Physics of the National Academy of Sciences of
Ukraine, 14-b Metrolohichna str. Kiev, 03143, Ukraine.}

\begin{abstract}
We investigate the eigen oscillations of internal degrees of freedom (Higgs mode and Goldstone mode) of two-band superconductors using the extended time-dependent Ginzburg-Landau theory, formulated in a work Grigorishin (2021) \cite{grig2}, for the case of two coupled order parameters by both the internal proximity effect and the drag effect. It is demonstrated, that the Goldstone mode splits into two branches: common mode oscillations with the acoustic spectrum, which is absorbed by the gauge field, and anti-phase oscillations with the energy gap (mass) in the spectrum determined by the interband coupling, which can be associated with the Leggett mode. The Higgs mode splits into two branches also: a massive one, whose energy gap vanishes at the critical temperature $T_{c}$, another massive one, whose energy gap does not vanish at  $T_{c}$. It is demonstrated, that the second branch of the Higgs mode is nonphysical, and it, together with the Leggett mode, can be removed by the special choice of coefficient at the "drag" term in the Lagrangian. In the same time, such a choice leaves only one coherence length, thereby prohibiting so-called type-1.5 superconductors. We analyze experimental data about the Josephson effect between two-band superconductors. In particular, it is demonstrated, that the resonant enhancement of the DC current through a Josephson junction at a resonant bias voltage $V_{\mathrm{res}}$, when the Josephson frequency matches the frequency of some internal oscillation mode in two-band superconductors (banks), can be explained with the coupling between AC Josephson current and Higgs oscillations. Thus, explanation of this effect does not need the Leggett mode.
\end{abstract}



\maketitle

\section{Introduction}\label{intro}

In a previous work \cite{grig2} the extended time-dependent Ginzburg-Landau (ETDGL) theory has been formulated, which is a generalization of the Ginsburg-Landau (GL) theory for the nonstationary regimes: the damped eigen oscillations (including the relaxation) and the forced oscillations of the order parameter (OP) $\Psi(\mathbf{r},t)$ under the action of an external field. In this theory an action with Lorentz invariant Lagrangian for the complex scalar field $\Psi=|\Psi|e^{i\theta}$ and the gauge field $A^{\mu}=(\varphi, \mathbf{A})$ in some 4D Minkowski space $\{\upsilon t,\mathbf{r}\}$, where speed $\upsilon$ is determined with the dynamical properties of the system, has been proposed. At the same time, the dynamics of conduction electrons remains non-relativistic. Accounting of movement of the normal component, which is accompanied by friction, makes the theory be not Lorentz covariant.

The superconducting (SC) system has two types of collective excitations: with an energy gap (quasi-relativistic spectrum) $E^{2}=\widetilde{m}^{2}\upsilon^{4}+p^{2}\upsilon^{2}$ (where $\widetilde{m}$ is the mass of a Higgs boson, so that $\widetilde{m}\upsilon^{2}=2|\Delta|$) - Higgs mode, and with an acoustic (ultrarelativistic) spectrum $E=p\upsilon$ - Goldstone mode. The light speed $\upsilon$ is determined from the dynamical properties of the system, and it is much less than the vacuum light speed: $\upsilon=v_{F}/\sqrt{3}\ll c$ ($v_{F}$ is Fermi velocity). The Higgs mode is oscillations of modulus of the OP $|\Psi(t,\mathbf{r})|$, and it can be considered as sound in the gas of above-condensate quasiparticles (at $T\rightarrow T_{c}$). The propagation of a Higgs boson is not accompanied by charge transfer. It should be noted, that the free Higgs mode is unstable due to both strong damping of these oscillations at $T\rightarrow T_{c}$, so that the aperiodic relaxation takes place with decay into above-condensate quasiparticles, since $E(q)\geq 2|\Delta|$. The Goldstone mode is oscillations of the phase $\theta(\mathbf{r},t)$, and it is absorbed into the gauge field $A^{\mu}$ according to the Anderson-Higgs mechanism. The Goldstone oscillations cannot be accompanied by oscillations of the charge density, they generate the transverse field $\textrm{div}\mathbf{A}=0$ only and they are eddy currents (i.e. $\textrm{div}\mathbf{j}=0$), as result of the boundary conditions. From the gauge invariance of the Lagrangian it follows, that the superconductor is equivalent to a dielectric (in some effective sense) with the permittivity $\varepsilon=\frac{c^{2}}{\upsilon^{2}}$ only for the induced electric field $\mathbf{E}=-\frac{1}{c}\frac{\partial\mathbf{A}}{\partial t}$ (with frequencies $0<\hbar\omega<2|\Delta|$). Thus, the speed $\upsilon$ is the speed of light in SC medium, if there were no skin-effect and Meissner effect. At the same time, inside the superconductor the potential electric field is absent $\mathbf{E}=-\nabla\varphi=0$ as consequence of the boundary conditions. For the electrostatic field $\mathbf{E}=-\nabla\varphi$ the permittivity is $\varepsilon(\omega=0,\mathbf{q}=0)=\infty$ as in metals.

Two-band superconductors are a specific class of superconductors essentially differing in their properties from single-band superconductors. The main feature of these materials is the presence of two OP - "wave functions" $\Psi_{1}$ and $\Psi_{2}$ corresponding to the condensates of Cooper pairs in each band. In a bulk isotropic s-wave superconductor the GL free energy functional can be written as
\cite{grig1,asker7,asker1,asker2,asker3,asker4,asker5,doh,yerin1,aguir}:
\begin{eqnarray}\label{1.1}
    F&=&\int d^{3}r[\frac{\hbar^{2}}{4m_{1}}\left|\nabla\Psi_{1}\right|^{2}+\frac{\hbar^{2}}{4m_{2}}\left|\nabla\Psi_{2}\right|^{2}
    +\frac{\hbar^{2}}{4}\eta\left(\nabla\Psi_{1}\nabla\Psi_{2}^{+}+\nabla\Psi_{1}^{+}\nabla\Psi_{2}\right)\nonumber\\
    &+&a_{1}\left|\Psi_{1}\right|^{2}+a_{2}\left|\Psi_{2}\right|^{2}+\frac{b_{1}}{2}\left|\Psi_{1}\right|^{4}+\frac{b_{2}}{2}\left|\Psi_{2}\right|^{4}
    +\epsilon\left(\Psi_{1}^{+}\Psi_{2}+\Psi_{1}\Psi_{2}^{+}\right)],
\end{eqnarray}
where $m_{1,2}$ denote the effective mass of carriers in the corresponding band, the coefficients $a_{1,2}$ are given as $a_{i}=\gamma_{i}(T-T_{ci})$ where $\gamma_{i}$ are some constants, the coefficients $b_{1,2}$ are independent of temperature, the quantities $\epsilon$ and $\eta$ describe interband mixing of the two OP (proximity effect) and their gradients (drag effect), respectively. If we switch off the interband interactions $\epsilon=0$ and $\eta=0$, then we will have two independent superconductors with different critical temperatures $T_{c1}$ and $T_{c2}$ because the intraband interactions can be different. Thus, a two-band superconductor is understood as two single-band superconductors with the corresponding condensates of Cooper pairs $\Psi_{1}$ and $\Psi_{2}$ (so that densities of SC electrons are $n_{\mathrm{s}1}=2|\Psi_{1}|^{2}$ and $n_{\mathrm{s}2}=2|\Psi_{2}|^{2}$ accordingly), but these two condensates are coupled by both the internal proximity effect $\epsilon\left(\Psi_{1}^{+}\Psi_{2}+\Psi_{1}\Psi_{2}^{+}\right)$ and the "drag" effect $\eta\left(\nabla\Psi_{1}\nabla\Psi_{2}^{+}+\nabla\Psi_{1}^{+}\nabla\Psi_{2}\right)$. In presence of a magnetic potential $\textbf{A}$ the replacement $\nabla\rightarrow\nabla-\frac{i2e}{c\hbar}\textbf{A}$ must be done in the free energy functional (\ref{1.1}) for the gauge invariance. The magnetic response (penetration, critical fields etc.) has been considered in Ref.\cite{grig1,asker1,asker2,asker3,asker4,asker5,asker7}.

Minimization of the free energy functional with respect to the OP, if $\nabla\Psi_{1,2}=0$, gives
\begin{equation}\label{1.2}
\left\{\begin{array}{c}
  a_{1}\Psi_{1}+\epsilon\Psi_{2}+b_{1}\Psi_{1}^{3}=0 \\
  a_{2}\Psi_{2}+\epsilon\Psi_{1}+b_{2}\Psi_{2}^{3}=0 \\
\end{array}\right\},
\end{equation}
where the equilibrium values $\Psi_{1,2}$ are assumed to be real (i.e. the phases $\theta_{1,2}$ are $0$ or $\pi$) in absence of current and magnetic field in the case of a two-band superconductor (but not for three-band superconductors, where the equilibrium phase differences can be not only $0$ or $\pi$, but it can be $2\pi/3$ or $\pi/3,2\pi/3$ depending on the signs of the interband interactions $\epsilon_{ik}$ \cite{stanev1,stanev2,tanaka}, and chiral ground states triggered by the interband interaction occur \cite{tanaka}). Near the critical temperature $T_{c}$ we have $\Psi_{1,2}^{3}\rightarrow 0$, hence, we can find the critical temperature equating to zero the determinant of the linearized system (\ref{1.2}):
\begin{equation}\label{1.3}
a_{1}a_{2}-\epsilon^{2}=\gamma_{1}\gamma_{2}(T_{c}-T_{c1})(T_{c}-T_{c2})-\epsilon^{2}=0.
\end{equation}
Solving this equation, we find $T_{c}>T_{c1},T_{c2}$, moreover, the solution does not depend on the sign of $\epsilon$. The sign determines the equilibrium phase difference of the OP $|\Psi_{1}|e^{i\theta_{1}}$ and $|\Psi_{2}|e^{i\theta_{2}}$:
\begin{equation}\label{1.4}
    \begin{array}{cc}
      \cos(\theta_{1}-\theta_{2})=1 & \mathrm{if}\quad\epsilon<0  \\
      \cos(\theta_{1}-\theta_{2})=-1 & \mathrm{if}\quad\epsilon>0 \\
    \end{array},
\end{equation}
that follows from Eq.(\ref{1.2}). Then, in the linear approximation at $T\rightarrow T_{c}$ ($T>T_{c1},T_{c2}$) we have $\Psi_{2}=-\frac{a_{1}}{\epsilon}\Psi_{1}=-\mathrm{sgn}(\epsilon)\frac{a_{1}}{|\epsilon|}\Psi_{1}
=-\mathrm{sgn}(\epsilon)\sqrt{\frac{a_{1}}{a_{2}}}\Psi_{1}$. The case $\epsilon<0$ corresponds to an attractive interband interaction (for example, in $\mathrm{MgB}_{2}$, where $s^{++}$ wave symmetry occurs), the case $\epsilon>0$ corresponds to a repulsive interband interaction (for example, in iron-based superconductors, where $s^{+-}$ wave symmetry occurs) \cite{asker7}. Thus, in the absence of interband coupling we obtain two independent bands, in which the OP can exhibit any sign. In the converse situation, the signs of the OP are determined by the sign of the interband coupling, that has been demonstrated with numerical calculations in the minimal two-orbital model of iron based superconductors in \cite{kapcia1}.

\begin{figure}[h]
\includegraphics[width=8.5cm]{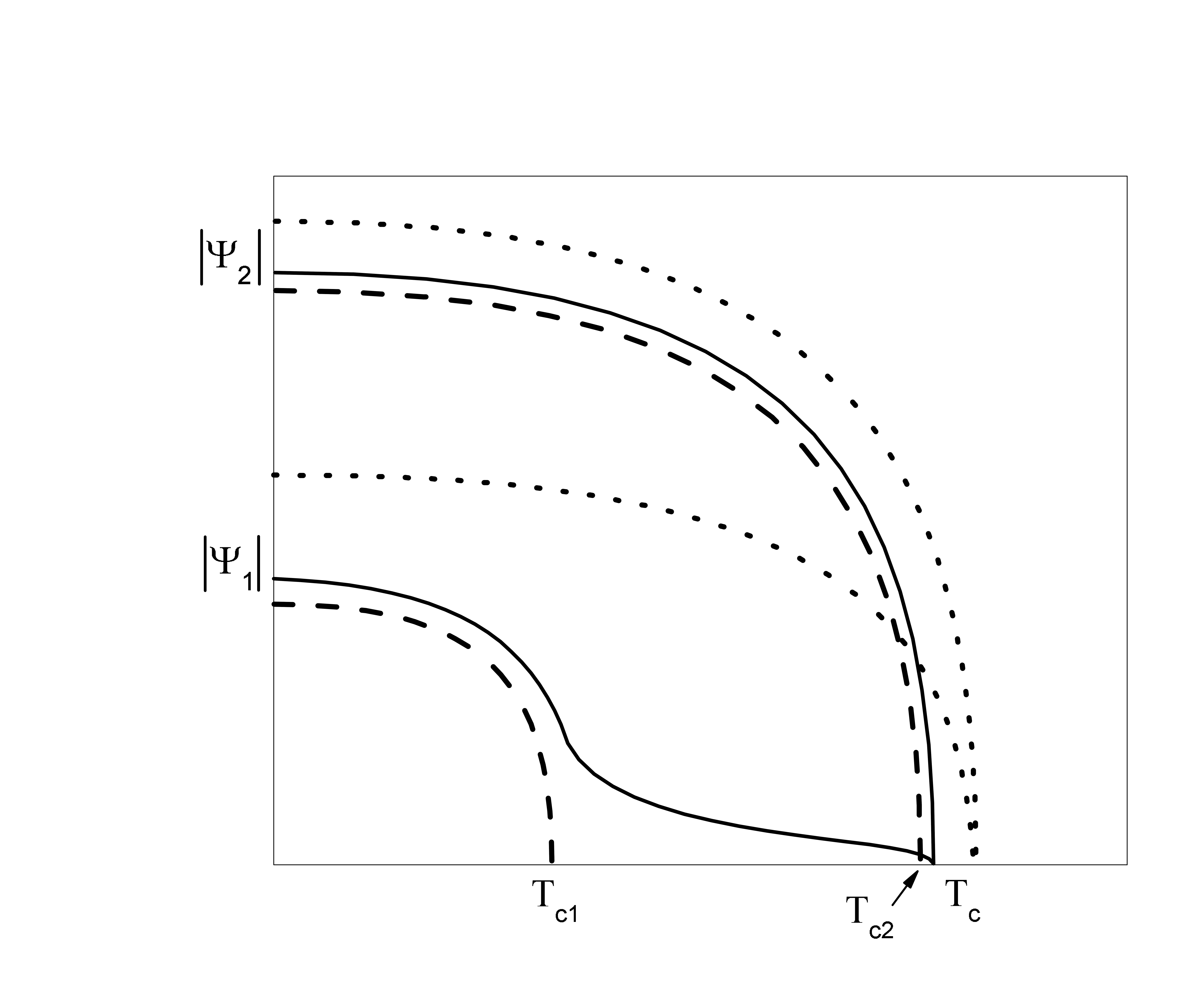}
\caption{The OP $\Psi_{1}(T)$ and $\Psi_{2}(T)$ as solutions of Eq.(\ref{1.2}), if the interband coupling is absent, i.e. $\epsilon=0$ (dash lines), and if the interband coupling is weak, i.e. $\epsilon\neq 0$, $|\epsilon|\ll |a_{1}(0)|$ (solid lines). The applying of the weak interband coupling washes out the smaller parameter $\Psi_{1}$ up to new critical temperature $T_{c}\gg T_{c1}$. As the coupling $|\epsilon|$ increases, $\Psi_{1}(T)$ and $\Psi_{2}(T)$ take forms shown with the dot lines. The effect on the larger parameter $\Psi_{2}$ is not so significant.}
\label{Fig1}
\end{figure}

The solutions of Eq.(\ref{1.2}) are illustrated in Fig.\ref{Fig1} for the case of strongly asymmetrical bands $T_{c1}\ll T_{c2}$. We can see, that the effect of interband coupling $\epsilon\neq 0$, even if the coupling is weak $|\epsilon|\ll |a_{1}(0)|$, is nonperturbative for the smaller OP $\Psi_{1}$: applying of the interband coupling washes out the smaller parameter up to the new critical temperature $T_{c}\gg T_{c1}$. At the same time, the effect on the larger parameter $\Psi_{2}$ is not so significant - applying of the interband coupling only slightly increases the critical temperature $T_{c}\gtrsim T_{c2}$. This property corresponds to numerical solutions of self-consistent equations for superconducting gaps $\Delta_{1}$ and $\Delta_{2}$ in two-band systems with both $s$-wave and $d$-wave symmetries \cite{nikol,litak1,litak2,kapcia2,gupta} and have been most clearly observed for $\mathrm{Mg}_{1-x}\mathrm{Al}_{x}\mathrm{B}_{2}$ and $\mathrm{LaO}_{1-x}\mathrm{F}_{x}\mathrm{FeAs}$ in Ref.\cite{kuzm} and $\mathrm{Mg}\mathrm{B}_{2}$ in Ref.\cite{pono1}. It is noteworthy, that the lower critical temperature $T_{c1}$ makes some mark in the temperature dependence of the specific heat $C(T)$: convexity, curvature and possible non-monotonicity near the temperature $T_{c1}$, which is demonstrated by numerical calculations in Ref.\cite{kapcia2} and has been observed experimentally for some iron-based superconductors $\mathrm{KFe}_{2}\mathrm{As}_{2}$ \cite{abdel}, $\mathrm{Lu}_{2}\mathrm{Fe}_{3}\mathrm{Si}_{5}$ \cite{nakajima} and for classical two-band superconductor $\mathrm{MgB}_{2}$ \cite{budko}.

In references \cite{litak1,litak2,ord1,ord2} it has been shown, that in a two-band superconductor there are two coherence lengths, which are not related to the concrete bands involved in the formation of SC state in a system with the interband interaction: one of the lengths  diverges at the critical temperature $\xi_{1}(T\rightarrow T_{c})\rightarrow\infty$, the second of them is a slightly varying function of temperature $\xi_{2}(T)\approx const$. At the same time, an isotropic two-band superconductor is characterized by a single magnetic penetration depth $\lambda(T)$ \cite{grig1,asker7,asker1,asker2,asker3,asker4,asker5}. Thus, we obtain two GL parameters: $\kappa_{1}=\lambda/\xi_{1}$ and $\kappa_{2}=\lambda/\xi_{2}$, which can be $\kappa_{1}<1/\sqrt{2}$ and $\kappa_{2}>1/\sqrt{2}$, that manifests about the new type of superconductivity – a novel "type-1.5 superconductor", contrary to type-I and type-II superconductors \cite{moshcal1,moshcal2,babaev1,babaev2,babaev3}.

In Ref.\cite{grig1} it has been shown, that the term of the drag effect $\eta\left(\nabla\Psi_{1}\nabla\Psi_{2}^{+}+\nabla\Psi_{1}^{+}\nabla\Psi_{2}\right)$ in the free energy functional of two-band superconductor plays an important role and the restrictions for the coefficient $\eta$ exist. If the coefficient is $\eta^{2}=\frac{1}{m_{1}m_{2}}$ and it’s sign is opposite to the sign of the coefficient in the term of the proximity effect $\epsilon\left(\Psi_{1}^{+}\Psi_{2}+\Psi_{1}\Psi_{2}^{+}\right)$, that is $\eta\epsilon<0$, then this leads to single coherence length $\xi$, which diverges at the critical temperature $\xi(T\rightarrow T_{c}) \rightarrow\infty$, and to single GL parameter. This quantity ensures  stability of SC state and the least possible free energy in this case. Other quantities of the coefficient or neglecting of the drag effect $\eta=0$ leads, at first, to the existence of two coherence lengths, where one of them diverges at the critical temperature, while the second length is finite at all temperatures. Secondly, it leads to the dynamical instability (suppressing of SC state, if the OP are spatial inhomogeneous) due to violation of the phase relations (\ref{1.4}). These results mean, that isotropic bulk type-1.5 superconductors are impossible. Using the results about the drag effect, it has been shown, that the free energy functional of a two-band superconductor can be reduced to GL functional for some effective single-band superconductor.

As has been demonstrated in Ref.\cite{grig2}, in two-band superconductors the Goldstone mode splits into two branches: common mode oscillations with acoustic spectrum, and the oscillations of the relative phase $\theta_{1}-\theta_{2}$ between two SC condensates with the energy gap in the spectrum determined by the interband coupling  - the Leggett mode (for symmetrical condensates we have the gap: $\sqrt{8|\epsilon|m\upsilon^{2}}$). The common mode oscillations are absorbed into the gauge field $A_{\mu}$ as in single-band superconductors, at the same time, the Leggett mode "survives" due to these oscillations are not accompanied by current. However, in this work, firstly, Higgs oscillations in two-band superconductors have not been investigated, secondly, effect of the drag term $\eta\left(\nabla\Psi_{1}\nabla\Psi_{2}^{+}+\nabla\Psi_{1}^{+}\nabla\Psi_{2}\right)$ has not been investigated either. At the same time, as we can see from Ref.\cite{grig1}, the drag term can have fundamental importance.

Proceeding from the aforesaid, we aim to obtain eigen oscillations of internal degrees of freedom (the Higgs mode and the Goldstone mode) of two-band superconductors using the ETDGL theory formulated in Ref.\cite{grig2} for the case of two coupled OP by both the internal proximity effect and the drag effect. Our paper is organized as follows. In Sect.\ref{Goldstone and Higgs} we generalize the two-band free energy (\ref{1.1}) to some action describing non-stationary regimes. Then, we obtain the spectra of both the Higgs mode and the Goldstone mode, which split into two branches each: common mode oscillations and anti-phase oscillations. It is demonstrated, that the second branch of the Higgs mode has a nonphysical property and it, together with the Leggett mode, can be removed with special choice of the coefficient $\eta$. Thus, we demonstrate, that the drag effect plays a fundamental role in the internal dynamics of two-band superconductors. In Sect.\ref{Josephson} we analyze experimental data about the Josephson effect between two-band superconductors. In particular, it is demonstrated, that the resonant enhancement of the DC current through a Josephson junction at a resonant bias voltage $V_{\mathrm{res}}$, when the Josephson frequency matches the frequency of some internal oscillation mode in two-band superconductors (banks), can be explained with the coupling between AC Josephson current and Higgs oscillations in two-band superconductors. Thus, explanation of this effect does not need the Leggett mode.

\section{Goldstone and Higgs oscillations in two-band superconductors}\label{Goldstone and Higgs}

\subsection{Ginzburg-Landau Lagrangian for two-band superconductors}

In the general case the OP $\Psi_{1,2}$ are both spatially inhomogeneous and they can change over time:
$\Psi_{1,2}=\Psi_{1,2}(\textbf{r},t)$. The OP are complex scalar fields, which are equivalent to two real fields each: modulus $\left|\Psi(\textbf{r},t)\right|$ and phase $\theta(\textbf{r},t)$ (the modulus-phase representation):
\begin{equation}\label{1.5}
    \Psi_{1}(\textbf{r},t)=\left|\Psi_{1}(\textbf{r},t)\right|e^{i\theta_{1}(\textbf{r},t)}, \quad\Psi_{2}(\textbf{r},t)=\left|\Psi_{2}(\textbf{r},t)\right|e^{i\theta_{2}(\textbf{r},t)}.
\end{equation}
For the stationary case $\Psi_{1,2}=\Psi_{1,2}(\textbf{r})$ the steady configuration of the field $\Psi_{1,2}(\textbf{r})$ minimizes the free energy functional (\ref{1.1}). However, for the nonstationary case $\Psi_{1,2}(\textbf{r},t)$ the minimization procedure makes no sense. According to the method described in \cite{grig2}, the parameter $t$ - the time can be turned into a coordinate $t\rightarrow\upsilon t$ in some 4D Minkowski space $\{\upsilon t,\textbf{r}\}$, where $\upsilon$ is an parameter of dimension of speed (as the light speed), which must be determined from the dynamical properties of the system. At the same time, the dynamics of conduction electrons remains non-relativistic. Then the two-component scalar fields $\Psi_{1,2}(\textbf{r},t)$ minimizes some action $S$ (as in the relativistic field theory \cite{sad}) in the Minkowski space:
\begin{equation}\label{1.6}
    S=\frac{1}{\upsilon}\int\mathcal{L}(\Psi_{1},\Psi_{2},\Psi^{+}_{1},\Psi^{+}_{2})\upsilon dtd^{3}r.
\end{equation}
The Lagrangian $\mathcal{L}$ is built by generalizing the density of free energy in Eq.(\ref{1.1}) to the "relativistic" invariant form by substitution of covariant and contravariant differential operators:
\begin{equation}\label{1.7}
\widetilde{\partial}_{\mu}\equiv\left(\frac{1}{\upsilon}\frac{\partial}{\partial t},\nabla\right),
\quad\widetilde{\partial}^{\mu}\equiv\left(\frac{1}{\upsilon}\frac{\partial}{\partial t},-\nabla\right),
\end{equation}
instead the gradient operators:
$\nabla\Psi\rightarrow\widetilde{\partial}_{\mu}\Psi,\quad\nabla\Psi^{+}\rightarrow\widetilde{\partial}^{\mu}\Psi^{+}$. Then the required Lagrangian takes the form:
\begin{eqnarray}\label{1.8}
    \mathcal{L}&=&\frac{\hbar^{2}}{4m_{1}}\widetilde{\partial}_{\mu}\Psi_{1}\widetilde{\partial}^{\mu}\Psi_{1}^{+}
    +\frac{\hbar^{2}}{4m_{2}}\widetilde{\partial}_{\mu}\Psi_{2}\widetilde{\partial}^{\mu}\Psi_{2}^{+}
    +\frac{\hbar^{2}}{4}\eta\left(\widetilde{\partial}_{\mu}\Psi_{1}\widetilde{\partial}^{\mu}\Psi_{2}^{+}
    +\widetilde{\partial}^{\mu}\Psi_{1}^{+}\widetilde{\partial}_{\mu}\Psi_{2}\right)\nonumber\\
    &-&a_{1}\left|\Psi_{1}\right|^{2}-\frac{b_{1}}{2}\left|\Psi_{1}\right|^{4}-a_{2}\left|\Psi_{2}\right|^{2}-\frac{b_{2}}{2}\left|\Psi_{2}\right|^{4}
    -\epsilon\left(\Psi_{1}^{+}\Psi_{2}+\Psi_{1}\Psi_{2}^{+}\right),
\end{eqnarray}
where the same speed $\upsilon$ is used for both $\Psi_{1}$ and $\Psi_{2}$ with the masses $m_{1}$ and $m_{2}$ accordingly (it should be noted the property $\widetilde{\partial}_{\mu}\Psi_{1}\widetilde{\partial}^{\mu}\Psi_{2}^{+}=\widetilde{\partial}^{\mu}\Psi_{1}\widetilde{\partial}_{\mu}\Psi_{2}^{+}$, etc.). The speed $\upsilon\sim\upsilon_{F1},\upsilon_{F2}$ (here, $\upsilon_{F1,2}$ are Fermi velocities in each band accordingly) plays role of the light speed in SC medium, and it will be found below. In the presence of el.-mag. field $A_{\mu}=(\varphi,-\textbf{A})$, the replacement $\partial_{\mu}\rightarrow\left(\partial_{\mu}+\frac{i2e}{c\hbar}\widetilde{A}_{\mu}\right)$
must be done in Lagrangian (\ref{1.8}) for a gauge invariance (here, $\widetilde{A}_{\mu}=(\frac{c}{\upsilon}\varphi,-\textbf{A})$).
The electromagnetic response has been considered in Ref.\cite{grig2}. In the present work we will consider only collective excitations, which are not accompanied by a current and a charge transfer (the Higgs mode, the Leggett mode etc.). Substituting representation (\ref{1.5}) in the Lagrangian (\ref{1.8}) we obtain:
\begin{eqnarray}\label{1.9}
    \mathcal{L}&=&\frac{\hbar^{2}}{4m_{1}}\widetilde{\partial}_{\mu}|\Psi_{1}|\widetilde{\partial}^{\mu}|\Psi_{1}|
    +\frac{\hbar^{2}}{4m_{1}}|\Psi_{1}|^{2}\widetilde{\partial}_{\mu}\theta_{1}\widetilde{\partial}^{\mu}\theta_{1}
    +\frac{\hbar^{2}}{4m_{2}}\widetilde{\partial}_{\mu}|\Psi_{2}|\widetilde{\partial}^{\mu}|\Psi_{2}|
    +\frac{\hbar^{2}}{4m_{2}}|\Psi_{2}|^{2}\widetilde{\partial}_{\mu}\theta_{2}\widetilde{\partial}^{\mu}\theta_{2}\nonumber\\
    &+&\frac{\hbar^{2}}{4}\eta\left(\widetilde{\partial}_{\mu}|\Psi_{1}|\widetilde{\partial}^{\mu}|\Psi_{2}|
    +\widetilde{\partial}_{\mu}|\Psi_{2}|\widetilde{\partial}^{\mu}|\Psi_{1}|\right)\cos\left(\theta_{1}-\theta_{2}\right)
    +\frac{\hbar^{2}}{4}\eta\left(\widetilde{\partial}_{\mu}\theta_{1}\widetilde{\partial}^{\mu}\theta_{2}
    +\widetilde{\partial}_{\mu}\theta_{2}\widetilde{\partial}^{\mu}\theta_{1}\right)|\Psi_{1}||\Psi_{2}|
    \cos\left(\theta_{1}-\theta_{2}\right)\nonumber\\
    &+&\frac{\hbar^{2}}{4}\eta\left(\widetilde{\partial}_{\mu}|\Psi_{1}|\widetilde{\partial}^{\mu}\theta_{2}
    +\widetilde{\partial}^{\mu}|\Psi_{1}|\widetilde{\partial}_{\mu}\theta_{2}\right)|\Psi_{2}|\sin\left(\theta_{1}-\theta_{2}\right)
    -\frac{\hbar^{2}}{4}\eta\left(\widetilde{\partial}_{\mu}|\Psi_{2}|\widetilde{\partial}^{\mu}\theta_{1}
    +\widetilde{\partial}^{\mu}|\Psi_{2}|\widetilde{\partial}_{\mu}\theta_{1}\right)|\Psi_{1}|\sin\left(\theta_{1}-\theta_{2}\right)\nonumber\\
    &-&a_{1}\left|\Psi_{1}\right|^{2}-\frac{b_{1}}{2}\left|\Psi_{1}\right|^{4}-a_{2}\left|\Psi_{2}\right|^{2}-\frac{b_{2}}{2}\left|\Psi_{2}\right|^{4}
    -2|\Psi_{1}||\Psi_{2}|\epsilon\cos\left(\theta_{1}-\theta_{2}\right),
\end{eqnarray}
where we have used the following properties, as $\widetilde{\partial}_{\mu}|\Psi_{1}|\widetilde{\partial}^{\mu}|\Psi_{2}|=\widetilde{\partial}_{\mu}|\Psi_{2}|\widetilde{\partial}^{\mu}|\Psi_{1}|$ and $\widetilde{\partial}_{\mu}|\Psi_{1}|\widetilde{\partial}^{\mu}\theta_{2}=\widetilde{\partial}^{\mu}|\Psi_{1}|\widetilde{\partial}_{\mu}\theta_{2}$. We will see, that modulus and phase variables can be separated in the linear approximation. Thus, the field coordinates $|\Psi_{1,2}(\mathbf{r},t)|$ and $\theta_{1,2}(\mathbf{r},t)$ are the normal coordinates, and their small oscillations are the normal oscillations.

\subsection{Goldstone oscillations}

Let us consider movement of the phases $\theta_{1,2}$. The corresponding Lagrange equations are
\begin{eqnarray}
  \widetilde{\partial}_{\mu}\frac{\partial\mathcal{L}}{\partial(\widetilde{\partial}_{\mu}\theta_{1})}
  -\frac{\partial\mathcal{L}}{\partial\theta_{1}}=0\Rightarrow&&
  \frac{\hbar^{2}}{4m_{1}}|\Psi_{1}|^{2}\widetilde{\partial}_{\mu}\widetilde{\partial}^{\mu}\theta_{1}+
  \frac{\hbar^{2}}{4}|\Psi_{1}||\Psi_{2}|\eta\cos\left(\theta_{1}-\theta_{2}\right)\widetilde{\partial}_{\mu}\widetilde{\partial}^{\mu}\theta_{2}
-\nonumber\\
&&\frac{\hbar^{2}}{4}\eta|\Psi_{1}|\sin\left(\theta_{1}-\theta_{2}\right)\widetilde{\partial}_{\mu}\widetilde{\partial}^{\mu}|\Psi_{2}|
  -|\Psi_{1}||\Psi_{2}|\epsilon\sin\left(\theta_{1}-\theta_{2}\right) = 0\label{2.2a}\\
\widetilde{\partial}_{\mu}\frac{\partial\mathcal{L}}{\partial(\widetilde{\partial}_{\mu}\theta_{2})}
  -\frac{\partial\mathcal{L}}{\partial\theta_{2}}=0\Rightarrow&&
  \frac{\hbar^{2}}{4m_{1}}|\Psi_{2}|^{2}\widetilde{\partial}_{\mu}\widetilde{\partial}^{\mu}\theta_{2}+
  \frac{\hbar^{2}}{4}|\Psi_{1}||\Psi_{2}|\eta\cos\left(\theta_{1}-\theta_{2}\right)\widetilde{\partial}_{\mu}\widetilde{\partial}^{\mu}\theta_{1}
+\nonumber\\
&&\frac{\hbar^{2}}{4}\eta|\Psi_{2}|\sin\left(\theta_{1}-\theta_{2}\right)\widetilde{\partial}_{\mu}\widetilde{\partial}^{\mu}|\Psi_{1}|
  +|\Psi_{1}||\Psi_{2}|\epsilon\sin\left(\theta_{1}-\theta_{2}\right) = 0\label{2.2b}
\end{eqnarray}
where we have omitted nonlinear terms, such as $\widetilde{\partial}_{\mu}\theta\widetilde{\partial}^{\mu}\theta$, $\widetilde{\partial}_{\mu}|\Psi|\widetilde{\partial}^{\mu}|\Psi|$, $\widetilde{\partial}_{\mu}\theta\widetilde{\partial}^{\mu}|\Psi|$. The phases can be written in the form of harmonic oscillations:
\begin{equation}\label{2.3}
    \begin{array}{c}
      \theta_{1}=\theta_{1}^{0}+Ae^{i(\mathbf{qr}-\omega t)}\equiv\theta_{1}^{0}+Ae^{-iq_{\mu}x^{\mu}} \\
      \theta_{2}=\theta_{2}^{0}+Be^{i(\mathbf{qr}-\omega t)}\equiv\theta_{2}^{0}+Be^{-iq_{\mu}x^{\mu}} \\
    \end{array},
\end{equation}
where $q_{\mu}=\left(\frac{\omega}{\upsilon},-\mathbf{q}\right)$, $x^{\mu}=\left(\upsilon t,\mathbf{r}\right)$, equilibrium phases $\theta_{1,2}^{0}$ satisfy the relation (\ref{1.4}), so that $\epsilon\cos\left(\theta_{1}^{0}-\theta_{2}^{0}\right)=-|\epsilon|$. Linearizing Eqs.(\ref{2.2a},\ref{2.2b}) by using relations $\cos\left(\theta_{1}-\theta_{2}\right)\approx-\epsilon/|\epsilon|$, $\epsilon\sin(\theta_{1}-\theta_{2})\approx-|\epsilon|\left(\theta_{1}-\theta_{2}-\theta_{1}^{0}+\theta_{2}^{0}\right)$ and by omitting the
nonlinear terms $\sin\left(\theta_{1}-\theta_{2}\right)\widetilde{\partial}_{\mu}\widetilde{\partial}^{\mu}|\Psi_{1,2}|$, we obtain the following linear equations:
\begin{eqnarray}
  \frac{\hbar^{2}}{4m_{1}}|\Psi_{1}|^{2}\widetilde{\partial}_{\mu}\widetilde{\partial}^{\mu}\theta_{1}
  -\frac{\hbar^{2}}{4}|\Psi_{1}||\Psi_{2}|\frac{\eta\epsilon}{|\epsilon|}\widetilde{\partial}_{\mu}\widetilde{\partial}^{\mu}\theta_{2}
  +|\Psi_{1}||\Psi_{2}|\epsilon|\left(\theta_{1}-\theta_{2}-\theta_{1}^{0}+\theta_{2}^{0}\right) &=& 0\label{2.4a}\\
  \frac{\hbar^{2}}{4m_{2}}|\Psi_{2}|^{2}\widetilde{\partial}_{\mu}\widetilde{\partial}^{\mu}\theta_{2}
  -\frac{\hbar^{2}}{4}|\Psi_{1}||\Psi_{2}|\frac{\eta\epsilon}{|\epsilon|}\widetilde{\partial}_{\mu}\widetilde{\partial}^{\mu}\theta_{1}
  -|\Psi_{1}||\Psi_{2}||\epsilon|\left(\theta_{1}-\theta_{2}-\theta_{1}^{0}+\theta_{2}^{0}\right) &=& 0.\label{2.4b}
\end{eqnarray}
We can see, that oscillations of the phases $\theta_{1,2}$ are separated from movement of the modules $|\Psi_{1,2}|$ in a linear approximation. Substituting the phases (\ref{2.3}) in Eqs.(\ref{2.4a},\ref{2.4b}) we obtain equations for the amplitudes $A$ and $B$:
\begin{equation}\label{2.5}
    \begin{array}{c}
      A\left(|\epsilon|-q_{\mu}q^{\mu}\frac{\hbar^{2}}{4m_{1}}\frac{|\Psi_{1}|}{|\Psi_{2}|}\right)
  +B\left(-|\epsilon|+q_{\mu}q^{\mu}\frac{\hbar^{2}}{4}\frac{\eta\epsilon}{|\epsilon|}\right)=0 \\
      A\left(-|\epsilon|+q_{\mu}q^{\mu}\frac{\hbar^{2}}{4}\frac{\eta\epsilon}{|\epsilon|}\right)
  +B\left(|\epsilon|-q_{\mu}q^{\mu}\frac{\hbar^{2}}{4m_{2}}\frac{|\Psi_{2}|}{|\Psi_{1}|}\right)=0
    \end{array}.
\end{equation}
Equating to zero the determinant of the system (\ref{2.5}), we find a dispersion equation:
\begin{equation}\label{2.6}
  \left(q_{\mu}q^{\mu}\right)^{2}\frac{\hbar^{2}}{4}\left[\frac{1}{m_{1}m_{2}}-\eta^{2}\right]
  =\left(q_{\mu}q^{\mu}\right)|\epsilon|
  \left[\frac{m_{1}|\Psi_{2}|^{2}+m_{2}|\Psi_{1}|^{2}}{m_{1}m_{2}|\Psi_{1}||\Psi_{2}|}-2\frac{\eta\epsilon}{|\epsilon|}\right].
\end{equation}
From where we can see, that one of dispersion relations is
\begin{equation}\label{2.7}
    q_{\mu}q^{\mu}=0\Rightarrow\omega^{2}=q^{2}\upsilon^{2},
\end{equation}
wherein $A=B$, thus this mode is common mode oscillations, as the Goldstone mode in single-band superconductors. There is another oscillation mode with spectrum
\begin{equation}\label{2.8}
  q_{\mu}q^{\mu}=\frac{4|\epsilon|}{\hbar^{2}}\left[\frac{1}{m_{1}m_{2}}-\eta^{2}\right]^{-1}
  \left[\frac{m_{1}|\Psi_{2}|^{2}+m_{2}|\Psi_{1}|^{2}}{m_{1}m_{2}|\Psi_{1}||\Psi_{2}|}-2\frac{\eta\epsilon}{|\epsilon|}\right],
\end{equation}
wherein
\begin{equation}\label{2.9}
    \frac{A}{B}=-\frac{m_{1}}{m_{2}}\frac{|\Psi_{2}|^{2}}{|\Psi_{1}|^{2}}\frac{1-m_{2}\frac{\eta\epsilon}{|\epsilon|}\frac{|\Psi_{1}|}{|\Psi_{2}|}}
    {1-m_{1}\frac{\eta\epsilon}{|\epsilon|}\frac{|\Psi_{2}|}{|\Psi_{1}|}},
\end{equation}
at that for symmetrical bands $m_{1}=m_{2},|\Psi_{1}|=|\Psi_{2}|$ we have $\frac{A}{B}=-1$. If we suppose the drag effect is absent: $\eta=0$, then
\begin{equation}\label{2.10}
    (\hbar\omega)^{2}=4|\epsilon|\frac{|\Psi_{1}|^{2}m_{2}+|\Psi_{2}|^{2}m_{1}}{|\Psi_{1}||\Psi_{2}|}\upsilon^{2}+(\hbar q)^{2}\upsilon^{2},
\end{equation}
wherein
\begin{equation}\label{2.11}
    \frac{A}{B}=-\frac{m_{1}}{m_{2}}\frac{|\Psi_{2}|^{2}}{|\Psi_{1}|^{2}}.
\end{equation}
For symmetrical bands $m_{1}=m_{2}\equiv m$ and $|\Psi_{1}|=|\Psi_{2}|$ we obtain
\begin{equation}\label{2.12}
    (\hbar\omega)^{2}=8|\epsilon|m\upsilon^{2}+(\hbar q)^{2}\upsilon^{2},\quad A=-B,
\end{equation}
that corresponds to results of Ref.\cite{grig2}. Thus, in two-band superconductors the Goldstone mode splits into two branches: common mode oscillations, where $\nabla\theta_{1}=\nabla\theta_{2}$, with acoustic spectrum - Eq.(\ref{2.7}), and the oscillations of the relative phase $\theta_{1}-\theta_{2}$ between two SC condensates (for symmetrical condensates we have $\nabla\theta_{1}=-\nabla\theta_{2}$) with the energy gap in spectrum determined by the interband coupling - Eqs.(\ref{2.8},\ref{2.10},\ref{2.12}), which can be identified as a Leggett mode \cite{legg,shar,yanag}. It should be noted, as we can see from Eqs.(\ref{2.8},\ref{2.10}), $\omega$ is not proportional to $|\Psi_{1}||\Psi_{2}|\propto|\Delta_{1}||\Delta_{2}|$. Moreover, $q_{\mu}q^{\mu}(T=T_{c})\propto |\epsilon|m\upsilon^{2}\neq 0$. It should be noted, that at $T>T_{c}$ the Goldstone excitations make no sense (as the light speed $\upsilon$ and the permittivity $c^{2}/\upsilon^{2}$ \cite{grig2}), so far as the phase $\theta$ of the OP makes no sense, since $\langle\Psi\rangle=0$, although $\langle|\Psi|^{2}\rangle\neq 0$ occurs due to fluctuations \cite{tinh}. \emph{If we suppose $\eta^{2}=\frac{1}{m_{1}m_{2}},\quad\eta\epsilon<0$, then from} Eq.(\ref{2.6}) \emph{we can see, that the Leggett mode is absent, and the common mode oscillations with spectrum} (\ref{2.7}) \emph{remains only}.

In a two band superconductor a current (flow) takes the following form \cite{grig1}:
\begin{eqnarray}\label{2.13}
\mathbf{j}&=&e\hbar\left[\frac{|\Psi_{1}|^{2}}{m_{1}}\nabla\theta_{1}+\eta|\Psi_{1}||\Psi_{2}|
  \left(\nabla\theta_{1}+\nabla\theta_{2}\right)\cos(\theta_{1}-\theta_{2})+\frac{|\Psi_{2}|^{2}}{m_{2}}\nabla\theta_{2}\right]\nonumber\\
&=&ie^{i(\mathbf{qr}-\omega t)}e\hbar\left[\frac{|\Psi_{1}|^{2}}{m_{1}}\left(1-m_{1}\frac{\eta\epsilon}{|\epsilon|} \frac{|\Psi_{2}|}{|\Psi_{1}|}\right)A+\frac{|\Psi_{2}|^{2}}{m_{2}}\left(1-m_{2}\frac{\eta\epsilon}{|\epsilon|} \frac{|\Psi_{1}|}{|\Psi_{2}|}\right)B\right]\mathbf{q},
\end{eqnarray}
from where we can see, that the Goldstone mode (\ref{2.7}) (where $A=B$) is accompanied by the current, therefore the gauge field $\widetilde{A}_{\mu}$ absorbs the Goldstone bosons $\theta_{1,2}$, as in single-band superconductors, i.e. the Anderson-Higgs mechanism takes place \cite{grig2}. \emph{For the Leggett mode (where $A/B$ is determined with Eq.(\ref{2.9})) we obtain $\mathbf{j}=0$, therefore such oscillations "survive" and can be observed.}

\subsection{Higgs oscillations}

In the previous subsection we could see, that oscillations of the phases $\theta_{1,2}$ are separated from movement of the modules $|\Psi_{1,2}|$ in the linear approximation. Therefore, let us consider only movement of the modules (that is, assuming $\theta_{1}=\theta_{1}^{0}$ and $\theta_{2}=\theta_{2}^{0}$), then Lagrangian (\ref{1.9}) takes the form:
\begin{eqnarray}\label{3.1}
    \mathcal{L}&=&\frac{\hbar^{2}}{4m_{1}}\widetilde{\partial}_{\mu}|\Psi_{1}|\widetilde{\partial}^{\mu}|\Psi_{1}|
    +\frac{\hbar^{2}}{4m_{2}}\widetilde{\partial}_{\mu}|\Psi_{2}|\widetilde{\partial}^{\mu}|\Psi_{2}|
    -\frac{\hbar^{2}}{4}\frac{\eta\epsilon}{|\epsilon|}\left(\widetilde{\partial}_{\mu}|\Psi_{1}|\widetilde{\partial}^{\mu}|\Psi_{2}|
    +\widetilde{\partial}_{\mu}|\Psi_{2}|\widetilde{\partial}^{\mu}|\Psi_{1}|\right)\nonumber\\
    &-&a_{1}\left|\Psi_{1}\right|^{2}-\frac{b_{1}}{2}\left|\Psi_{1}\right|^{4}-a_{2}\left|\Psi_{2}\right|^{2}-\frac{b_{2}}{2}\left|\Psi_{2}\right|^{4}
    +2|\Psi_{1}||\Psi_{2}||\epsilon|.
\end{eqnarray}
At $T<T_{c}$ we can consider small variations of modules of OP from their equilibrium values: $|\Psi_{1,2}|=\Psi_{01,02}+\phi_{1,2}$, where $|\phi_{1,2}|\ll\Psi_{01,02}$. Then, $|\Psi|^{2}\approx\Psi_{0}^{2}+2\Psi_{0}\phi+\phi^{2}$, $|\Psi|^{4}\approx\Psi_{0}^{4}+4\Psi_{0}^{3}\phi+6\Psi_{0}^{2}\phi^{2}$, $|\Psi_{1}||\Psi_{2}|\approx\Psi_{01}\Psi_{02}+\Psi_{01}\phi_{2}+\Psi_{02}\phi_{1}+\phi_{1}\phi_{2}$, and Lagrangian (\ref{3.1}) takes the form:
\begin{eqnarray}\label{3.2}
    \mathcal{L}&=&\frac{\hbar^{2}}{4m_{1}}\widetilde{\partial}_{\mu}\phi_{1}\widetilde{\partial}^{\mu}\phi_{1}
    +\frac{\hbar^{2}}{4m_{2}}\widetilde{\partial}_{\mu}\phi_{2}\widetilde{\partial}^{\mu}\phi_{2}
    -\frac{\hbar^{2}}{4}\frac{\eta\epsilon}{|\epsilon|}\left(\widetilde{\partial}_{\mu}\phi_{1}\widetilde{\partial}^{\mu}\phi_{2}
    +\widetilde{\partial}_{\mu}\phi_{2}\widetilde{\partial}^{\mu}\phi_{1}\right)\nonumber\\
    &-&\phi_{1}^{2}\left(a_{1}+3b_{1}\Psi_{01}^{2}\right)-\phi_{2}^{2}\left(a_{2}+3b_{2}\Psi_{02}^{2}\right)+2|\epsilon|\phi_{1}\phi_{2}\nonumber\\
    &+&2\phi_{1}\left(|\epsilon|\Psi_{02}-a_{1}\Psi_{01}-b_{1}\Psi_{01}^{3}\right)
    +2\phi_{2}\left(|\epsilon|\Psi_{01}-a_{2}\Psi_{02}-b_{2}\Psi_{02}^{3}\right)\nonumber\\
    &-&a_{1}\Psi_{01}^{2}-\frac{b_{1}}{2}\Psi_{01}^{4}-a_{2}\Psi_{02}^{2}-\frac{b_{2}}{2}\Psi_{02}^{4}+2\Psi_{01}\Psi_{02}|\epsilon|.
\end{eqnarray}
The last five terms can be omitted as a constant. The terms at $\phi_{1}$ and $\phi_{2}$ should be zero:
\begin{equation}\label{3.3}
  \left\{\begin{array}{c}
  a_{1}\Psi_{01}-|\epsilon|\Psi_{02}+b_{1}\Psi_{01}^{3}=0 \\
  a_{2}\Psi_{02}-|\epsilon|\Psi_{01}+b_{2}\Psi_{02}^{3}=0 \\
\end{array}\right\},
\end{equation}
that corresponds to Eq.(\ref{1.2}). At $T>T_{c1},T_{c2}$ we have $a_{1,2}>0$ and $\epsilon^{2}-a_{1}(T_{c})a_{2}(T_{c})=0$, at $T<T_{c1},T_{c2}$ we have $a_{1,2}<0$. For the case of \emph{the weak interband coupling} $\epsilon^{2}\ll a_{1}a_{2}$, at $T\ll T_{c1},T_{c2}$ it is not difficult to obtain from Eq.(\ref{3.3}):
\begin{equation}\label{3.4a}
\begin{array}{c}
 \Psi_{01}=\sqrt{\frac{|a_{1}|}{b_{1}}}
\left(1+\frac{|\epsilon|}{\sqrt{|a_{1}||a_{2}|}}\sqrt{\frac{b_{1}}{b_{2}}}\frac{|a_{2}|}{|a_{1}|}\right)\approx\sqrt{\frac{|a_{1}|}{b_{1}}}\\
 \Psi_{02}=\sqrt{\frac{|a_{2}|}{b_{2}}}
\left(1+\frac{|\epsilon|}{\sqrt{|a_{2}||a_{1}|}}\sqrt{\frac{b_{2}}{b_{1}}}\frac{|a_{1}|}{|a_{2}|}\right)\approx\sqrt{\frac{|a_{2}|}{b_{2}}}\\
\end{array}.
\end{equation}
That is, the effect of the weak interband coupling on the both OP $\Psi_{1,2}$ at $T=0$ is not significant, and it can be described as perturbation. Let, for example, $T_{c1}\ll T_{c2}$, then from Eq.(\ref{3.3}) we can obtain:
\begin{equation}\label{3.4e}
\begin{array}{ccc}
 \Psi_{01}^{3}=|\epsilon|\sqrt{\frac{|a_{2}|}{b_{2}b_{1}^{2}}}& \mathrm{at} & T=T_{c1}\\
 \Psi_{02}^{2}=\frac{\epsilon^{2}}{a_{1}b_{2}}& \mathrm{at} & T=T_{c2}\\
\end{array}.
\end{equation}
For symmetrical bands, i.e. $a_{1}=a_{2}$, $b_{1}=b_{2}\equiv b$, $\Psi_{01}=\Psi_{02}\equiv\Psi_{0}$, at $T=T_{c1}=T_{c2}$  we have
\begin{equation}\label{3.4d}
\Psi_{0}^{2}=\frac{|\epsilon|}{b}.
\end{equation}
At $T\rightarrow T_{c}$ we have $\Psi_{01,02}\rightarrow 0$, then it is not difficult to obtain from Eq.(\ref{3.3}):
\begin{equation}\label{3.4f}
 \Psi_{01}^{2}=\frac{\epsilon^{2}\left(\epsilon^{2}-a_{1}a_{2}\right)}{\epsilon^{2}a_{1}b_{2}+b_{1}a_{2}^{3}},\quad
 \Psi_{02}^{2}=\frac{\epsilon^{2}\left(\epsilon^{2}-a_{1}a_{2}\right)}{\epsilon^{2}a_{2}b_{1}+b_{2}a_{1}^{3}}.
\end{equation}
Thus, at high temperatures $T\gtrsim T_{c1},T_{c2}$, the values of $\Psi_{01,02}$ are determined by the interband coupling $\epsilon$, so that, if $\epsilon=0$, then $\Psi_{01,02}=0$.

Let us introduce the following notes:
\begin{equation}\label{3.4b}
  \alpha_{1}\equiv a_{1}+3b_{1}\Psi_{01}^{2}, \quad \alpha_{2}\equiv a_{2}+3b_{2}\Psi_{02}^{2},
\end{equation}
then,
\begin{equation}\label{3.4c}
\begin{array}{ccc}
  \alpha_{1,2}= a_{1,2}>0 & \mathrm{at} & T=T_{c} \\
  \alpha_{1,2}=-2a_{1,2}=2|a_{1,2}| & \mathrm{at} & T\ll T_{c1},T_{c2}\\
\end{array}.
\end{equation}
The second formula is correct only if the weak interband coupling $\epsilon^{2}\ll a_{1}a_{2}$ takes place. The Lagrange equations for the Lagrangian (\ref{3.2}) are:
\begin{eqnarray}
  \frac{\hbar^{2}}{4m_{1}}\widetilde{\partial}_{\mu}\widetilde{\partial}^{\mu}\phi_{1}
  -\frac{\hbar^{2}}{4}\frac{\eta\epsilon}{|\epsilon|}\widetilde{\partial}_{\mu}\widetilde{\partial}^{\mu}\phi_{2}
  +\alpha_{1}\phi_{1}-|\epsilon|\phi_{2} &=& 0\label{3.5a}\\
 \frac{\hbar^{2}}{4m_{2}}\widetilde{\partial}_{\mu}\widetilde{\partial}^{\mu}\phi_{2}
  -\frac{\hbar^{2}}{4}\frac{\eta\epsilon}{|\epsilon|}\widetilde{\partial}_{\mu}\widetilde{\partial}^{\mu}\phi_{1}
  +\alpha_{2}\phi_{2}-|\epsilon|\phi_{1} &=& 0.\label{3.5b}
\end{eqnarray}
The fields $\phi_{1,2}$ can be written in the form of harmonic oscillations: $\phi_{1}=Ae^{-iq_{\mu}x^{\mu}}$, $\phi_{2}=Be^{-iq_{\mu}x^{\mu}}$, where $q_{\mu}x^{\mu}=\omega t-\mathbf{qr}$. Substituting them in Eqs.(\ref{3.5a},\ref{3.5b}) we obtain equations for the amplitudes $A$ and $B$:
\begin{eqnarray}\label{3.6}
   \begin{array}{c}
   A\left(\alpha_{1}-q_{\mu}q^{\mu}\frac{\hbar^{2}}{4m_{1}}\right)
  +B\left(-|\epsilon|+q_{\mu}q^{\mu}\frac{\hbar^{2}}{4}\frac{\eta\epsilon}{|\epsilon|}\right)=0 \\
  A\left(-|\epsilon|+q_{\mu}q^{\mu}\frac{\hbar^{2}}{4}\frac{\eta\epsilon}{|\epsilon|}\right)
  +B\left(\alpha_{2}-q_{\mu}q^{\mu}\frac{\hbar^{2}}{4m_{2}}\right)=0
\end{array}.
\end{eqnarray}
Equating to zero the determinant of the system (\ref{3.6}), we find a dispersion equation:
\begin{equation}\label{3.7}
  \left(q_{\mu}q^{\mu}\right)^{2}\frac{\hbar^{4}}{16}\left[\frac{1}{m_{1}m_{2}}-\eta^{2}\right]
  -\left(q_{\mu}q^{\mu}\right)\frac{\hbar^{2}}{4}
  \left[\frac{\alpha_{1}}{m_{2}}+\frac{\alpha_{2}}{m_{1}}-2\eta\epsilon\right]+\alpha_{1}\alpha_{2}-\epsilon^{2}=0.
\end{equation}
From where we obtain the following dispersion relations:
\begin{equation}\label{3.8}
  q_{\mu}q^{\mu}=\frac{2}{\hbar^{2}}\left[\frac{1}{m_{1}m_{2}}-\eta^{2}\right]^{-1}
  \left[\left(\frac{\alpha_{1}}{m_{2}}+\frac{\alpha_{2}}{m_{1}}-2\eta\epsilon\right)\pm\sqrt{\mathfrak{D}}\right],
\end{equation}
where
\begin{equation}\label{3.9}
  \mathfrak{D}=\left(\frac{\alpha_{1}}{m_{2}}+\frac{\alpha_{2}}{m_{1}}-2\eta\epsilon\right)^{2}
  -4\left[\frac{1}{m_{1}m_{2}}-\eta^{2}\right]\left(\alpha_{1}\alpha_{2}-\epsilon^{2}\right).
\end{equation}
Thus, in two-band superconductors the Higgs mode splits into two branches. Let us consider each mode at $T=T_{c}$, that is $a_{1}a_{2}=\epsilon^{2}$ (however, we must apply for $\alpha_{1}\alpha_{2}$ more accurate expressions than Eq.(\ref{3.4c}) using Eq.(\ref{3.4f},\ref{3.4b}), so that $\alpha_{1}\alpha_{2}\approx a_{1}a_{2}+3a_{1}b_{2}\Psi_{02}^{2}+3a_{2}b_{1}\Psi_{01}^{2}$):
\begin{eqnarray}
  q_{\mu}q^{\mu} &=& \frac{4}{\hbar^{2}}\frac{f(T_{c})(\epsilon^{2}-a_{1}a_{2})}{\frac{a_{1}}{m_{2}}+\frac{a_{2}}{m_{1}}-2\eta\epsilon}=0 \label{3.10a}\\
  q_{\mu}q^{\mu} &=& \frac{4}{\hbar^{2}}\frac{\frac{a_{1}}{m_{2}}+\frac{a_{2}}{m_{1}}-2\eta\epsilon}{\frac{1}{m_{1}m_{2}}-\eta^{2}}\neq 0, \label{3.10b}
\end{eqnarray}
where $f(T_{c})$ is some finite dimensionless value. Obviously, the coefficient $\eta$ must be such, that $\frac{a_{1}}{m_{2}}+\frac{a_{2}}{m_{1}}-2\eta\epsilon>0$ and $\eta^{2}<\frac{1}{m_{1}m_{2}}$. We can see, that for the first mode (\ref{3.10a}) the energy gap (the mass of a Higgs boson) vanishes at the critical temperature, as in single-band superconductors. At the same time, the energy gap of the second mode (\ref{3.10b}) does not vanish at the critical temperature. So, in a case of symmetrical bands $m_{1}=m_{2}\equiv m$, $a_{1}=a_{2}\equiv a$ at $T=T_{c}$ (then $a(T_{c})=|\epsilon|$) and supposing, that the drag effect is absent $\eta=0$ we obtain:
\begin{equation}\label{3.10c}
    (\hbar\omega)^{2}=8|\epsilon|m\upsilon^{2}+(\hbar q)^{2}\upsilon^{2},
\end{equation}
which coincides with energy of the Leggett mode (\ref{2.12}). If $\eta^{2}=\frac{1}{m_{1}m_{2}},\quad\eta\epsilon<0$, then only the single mode takes place:
\begin{equation}\label{3.11}
  q_{\mu}q^{\mu}=\frac{4}{\hbar^{2}}
  \frac{\alpha_{1}\alpha_{2}-\epsilon^{2}}{\frac{\alpha_{1}}{m_{2}}+\frac{\alpha_{2}}{m_{1}}+\frac{2|\epsilon|}{\sqrt{m_{1}m_{2}}}},
\end{equation}
whose energy gap vanishes at $T=T_{c}$, as for the mode (\ref{3.10a}). In Ref.\cite{grig2} it has been demonstrated how the energy gap $\hbar\omega_{0}$ is related to coherence length $\xi$: $\xi^{2}=\frac{2\upsilon^{2}}{\omega_{0}^{2}}$ (or from the uncertainty principle: $\hbar\omega_{0}\frac{\xi}{\upsilon}\sim\hbar\Rightarrow\xi\sim\frac{\upsilon}{\omega_{0}}$, since the energy of the Higgs mode plays role of the uncertainty of energy in a superconductor). Thus, we obtain the coherence lengths accordingly to the branches (\ref{3.10a},\ref{3.10b}) at $T=T_{c}$:
\begin{eqnarray}
  \xi^{2} &=& \frac{\hbar^{2}}{2}\frac{\frac{a_{1}}{m_{2}}+\frac{a_{2}}{m_{1}}-2\eta\epsilon}{f(T_{c})\left|\epsilon^{2}-a_{1}a_{2}\right|}=\infty \label{3.11a}\\
  \xi^{2} &=& \frac{\hbar^{2}}{2}\frac{\frac{1}{m_{1}m_{2}}-\eta^{2}}{\frac{a_{1}}{m_{2}}+\frac{a_{2}}{m_{1}}-2\eta\epsilon}< \infty. \label{3.11b}
\end{eqnarray}
We can see, that the first coherence length diverges at $T=T_{c}$. On the contrary, the second length remains finite and it varies little with temperature. These length scales are not related to the concrete bands involved in the formation of the superconducting ordering in a system with interband interaction. This result corresponds to the results in Ref.\cite{litak1,litak2,ord1,ord2} obtained with a microscopic approach, however they suggest, that the intergradient interaction is absent (i.e. $\eta=0$), and corresponds to the results in Ref.\cite{grig1} obtained with a phenomenological approach.

Let us consider the ratio of amplitudes for both branches of the spectrum. From Eq.(\ref{3.6}) we have:
\begin{equation}\label{3.12}
    \frac{A}{B}=\frac{|\epsilon|-q_{\mu}q^{\mu}\frac{\hbar^{2}}{4}\frac{\eta\epsilon}{|\epsilon|}}
    {\alpha_{1}-q_{\mu}q^{\mu}\frac{\hbar^{2}}{4m_{1}}}\equiv
    \frac{\alpha_{2}-q_{\mu}q^{\mu}\frac{\hbar^{2}}
    {4m_{2}}}{|\epsilon|-q_{\mu}q^{\mu}\frac{\hbar^{2}}{4}\frac{\eta\epsilon}{|\epsilon|}}.
\end{equation}
At $T\rightarrow T_{c}$, using Eqs.(\ref{3.10a},\ref{3.10b}), we obtain:
\begin{eqnarray}
  q_{\mu}q^{\mu}\rightarrow 0 &\Rightarrow& \frac{A}{B}=\frac{|\epsilon|}{a_{1}}
  =\frac{a_{2}}{|\epsilon|}=\sqrt{\frac{a_{2}}{a_{1}}}>0\label{3.13a}\\
  q_{\mu}q^{\mu}\rightarrow \frac{4}{\hbar^{2}}\frac{\frac{a_{1}}{m_{2}}+\frac{a_{2}}{m_{1}}}{\frac{1}{m_{1}m_{2}}}
  &\Rightarrow& \frac{A}{B}=-\frac{|\epsilon|}{a_{2}}=-\frac{a_{1}}{|\epsilon|}=-\sqrt{\frac{a_{1}}{a_{2}}}<0,\label{3.13b}
\end{eqnarray}
where in Eq.(\ref{3.13b}) we suppose $\eta=0$ for simplicity. Thus, for the mode (\ref{3.10a}) (or for the mode (\ref{3.11}), i.e. when $\eta^{2}=\frac{1}{m_{1}m_{2}},\quad\eta\epsilon<0$), oscillations of $|\Psi_{1}|$ and $|\Psi_{2}|$ occur in phase. For the mode (\ref{3.10b}),  oscillations of $|\Psi_{1}|$ and $|\Psi_{2}|$ occur in anti-phase.

The reason for the difference in energies between the common mode and anti-phase Higgs oscillations is as follows. Let us rewrite Eq.(\ref{3.3}) for equilibrium values $\Psi_{01}$ and $\Psi_{02}$ in the form:
\begin{equation}\label{3.13c}
  \left\{\begin{array}{c}
  \Psi_{02}=\Psi_{01}(a_{1}+b_{1}\Psi_{01}^{2})/|\epsilon| \\
  \Psi_{01}=\Psi_{02}(a_{2}+b_{2}\Psi_{02}^{2})/|\epsilon| \\
\end{array}\right\}.
\end{equation}
From the first equation we can see, that increasing (decreasing) of $\Psi_{01}$ results in increasing (decreasing) of $\Psi_{02}$, then from the second equation we can see, that the increased (decreased) $\Psi_{02}$ causes increasing (decreasing) of $\Psi_{01}$. This increasing (decreasing) of $\Psi_{01}$ causes the increasing (decreasing) of $\Psi_{02}$ again, and so on. Thus, Eqs.(\ref{3.3}) describe a system with the positive feedback. Therefore, excitation of the common mode oscillations requires less energy, than if the condensates from different bands were independent. On the contrary, to excite the anti-phase oscillations ($\Psi_{01}$ increases, but $\Psi_{02}$ decreases and vice versa) we must do the work against the positive feedback, hence minimal energy of these oscillations is larger.

Thus, Higgs modes are oscillations of SC densities $n_{\mathrm{s}i}=2|\Psi_{i}|^{2}$, at the same time, the normal density must oscillate in anti-phase so, that $n=n_{\mathrm{s}}+n_{\mathrm{n}}=\mathrm{const}$ and $n_{\mathrm{s}}\mathbf{v}_{s}+n_{\mathrm{n}}\mathbf{v}_{n}=0$. To change SC density and, hence, the normal density, as least one Cooper pair must be broken, that is the energy of order of $2|\Delta|$ must be spent. So, in Ref.\cite{grig2} it has been demonstrated, that in single-band superconductors $q_{\mu}q^{\mu}=4|\Delta|^{2}$. \emph{Thus, excitation of any Higgs mode at $T=T_{c}$ does not require the energy consumption (for $\mathbf{q}=0$), since $|\Delta|=0$ (i.e. a Cooper pair has zero binding energy). However, we could see, that for the second branch} - Eqs.(\ref{3.10b},\ref{3.10c}) \emph{we have $q_{\mu}q^{\mu}(T_{c})\neq 0$, which is a nonphysical property}. It should be noted, that this argument is not correct in the case of superconductors with a pseudogap, where uncorrelated pairs can be formed at $T>T_{c}$ due to strong electron–electron interaction, but the phase coherence is possible only at $T<T_{c}$ \cite{dzum1,dzum2,dzum3,dzum4}.

Thus, we must suppose
\begin{equation}\label{3.14}
  \eta^{2}=\frac{1}{m_{1}m_{2}},\quad\eta\epsilon<0,
\end{equation}
then from Eq.(\ref{3.7}) we can see, that the anti-phase Higgs mode is absent, and the common mode oscillations with zero energy gap at $T=T_{c}$ (\ref{3.11}) remain only. Analogously, from Eq.(\ref{2.6}) we can see, that the Leggett mode is absent, and the common mode oscillations with gapless spectrum (\ref{2.7}) only remains. \emph{Thus, in two-band superconductors one Goldstone mode and one Higgs mode exist only, as in single-band superconductors, and the Leggett mode and the anti-phase Higgs mode are absent}. In the same time, the Goldstone mode is accompanied by current - Eq.(\ref{2.13}), therefore the gauge field $\widetilde{A}_{\mu}$ absorbs the Goldstone boson $\theta$, as in single-band superconductors, i.e. the Anderson-Higgs mechanism takes place \cite{grig2}. In addition, the coherence length (\ref{3.11a}) only remains, that prohibits type 1.5 superconductors and corresponds to the result of Ref.\cite{grig1}.

Let us consider regime of almost independent condensates in each bands. This means: 1) the temperature must be low, i.e. $T\ll T_{c1},T_{c2}$, 2) the weak interband coupling $\epsilon^{2}\ll a_{1}a_{2}$ must take place. Using Eqs.(\ref{3.4a},\ref{3.4c}), the energy gap $\hbar\omega_{0}$ ($\mathbf{q}=0$) of Higgs mode (\ref{3.11}) can be presented in the form:
\begin{equation}\label{3.15}
  (\hbar\omega_{0})^{2}=4\upsilon^{2}
  \frac{\alpha_{1}\alpha_{2}-\epsilon^{2}}{\frac{\alpha_{1}}{m_{2}}+\frac{\alpha_{2}}{m_{1}}+\frac{2|\epsilon|}{\sqrt{m_{1}m_{2}}}}
  \approx 4\upsilon^{2}\frac{4\sqrt{|a_{1}||a_{2}|b_{1}b_{2}}}{\frac{2|a_{1}|}{m_{2}}+\frac{2|a_{2}|}{m_{1}}}\Psi_{01}\Psi_{02}.
\end{equation}
Then the multiplier before $\Psi_{01}\Psi_{02}$ depends on temperature very weakly, and this energy is symmetrical with respect to the bands. Using relationship between the "wave function" of Cooper pairs $\Psi$ and the energy gap $\Delta$ \cite{grig2,sad1,levit}, which can be generalized for  two-band superconductors in the form:
\begin{equation}\label{3.16}
   \Psi_{1}=\frac{\left(14\zeta(3)n_{1}\right)^{1/2}}{4\pi T_{c1}}\Delta_{1},\quad\Psi_{2}=\frac{\left(14\zeta(3)n_{2}\right)^{1/2}}{4\pi T_{c2}}\Delta_{2},
\end{equation}
where $n_{1,2}=\frac{k_{F1,2}^{3}}{3\pi^{2}}$ are electron densities for each band. Then, we can see, that $(\hbar\omega_{0})^{2}\propto|\Delta_{1}||\Delta_{2}|$, and we can suppose:
\begin{equation}\label{3.17}
  (\hbar\omega_{0})^{2}=\chi\Delta_{01}\Delta_{02},
\end{equation}
where $\chi=\mathrm{const}$ (dimensionless) is such, that in superconductor with symmetrical $m_{1}=m_{2}$, $n_{1}=n_{2}$, $a_{1}=a_{2}$, $b_{1}=b_{2}$, $T_{c1}=T_{c2}$ $\Rightarrow$ $\Delta_{1}=\Delta_{2}$ and almost independent bands (i.e. $\epsilon^{2}\ll a_{1}a_{2}$ at $T\ll T_{c1},T_{c2}$), we should have $\upsilon=\frac{\upsilon_{F}}{\sqrt{6}}$, since in single-band superconductors we have $\upsilon=\frac{\upsilon_{F}}{\sqrt{3}}$ and we can determine the "dielectric permittivity" as $\varepsilon=\frac{c^{2}}{\upsilon^{2}}=\frac{c^{2}}{\upsilon_{F}^{2}/3}$ \cite{grig2}, then the "mixture" of two superconductors is equivalent to two parallel dielectrics (capacitors) with total permittivity $\varepsilon=\varepsilon_{1}+\varepsilon_{2}=\frac{2c^{2}}{\upsilon_{F}^{2}/3}$; hence, we obtain for the "mixture": $\upsilon=\frac{\upsilon_{F}}{\sqrt{6}}$. The coefficients $a_{1,2}$, $b_{1,2}$ are \cite{sad2}:
\begin{equation}\label{3.18}
  a_{1,2}=\frac{6\pi^{2}T_{c1,2}}{7\zeta(3)\varepsilon_{F1,2}}\left(T-T_{c1,2}\right),\quad b_{1,2}=\frac{6\pi^{2}T_{c1,2}}{7\zeta(3)\varepsilon_{F1,2}}\frac{T_{c1,2}}{n_{1,2}}.
\end{equation}
Substituting Eqs.(\ref{3.15},\ref{3.16},\ref{3.18}) in Eq.(\ref{3.17}) we obtain:
\begin{equation}\label{3.19}
  \upsilon^{2}=\frac{\chi}{12}\frac{\upsilon_{F1}^{2}T_{c2}(T_{c2}-T)+\upsilon_{F2}^{2}T_{c1}(T_{c1}-T)}
  {\sqrt{T_{c1}T_{c2}}\sqrt{(T_{c1}-T)(T_{c2}-T)}}.
\end{equation}
If we consider symmetrical bands, i.e. $\upsilon_{F1}=\upsilon_{F2}$ and $T_{c1}=T_{c2}$, then we obtain $\upsilon^{2}=\frac{\chi}{6}\upsilon_{F}^{2}\Rightarrow\chi=1$. It is not difficult to notice, that the dependence of $\upsilon$ on $T$ in Eq.(\ref{3.19}) is weak (at $T\ll T_{c1},T_{c2}$), hence we can suppose:
\begin{equation}\label{3.20}
  \upsilon^{2}\approx\frac{1}{12}\left(\frac{T_{c2}}{T_{c1}}\upsilon_{F1}^{2}+\frac{T_{c1}}{T_{c2}}\upsilon_{F2}^{2}\right).
\end{equation}
Strictly speaking, the relation $\Psi\propto\Delta$ is not valid at low temperatures $T\ll T_{c}$ both in pure superconductors (where $n_{s}(0)=n$) and in dirty superconductors (where $n_{s}(0)\propto|\Delta|$) \cite{sad1,levit} (here, $n_{s}=2|\Psi|^{2}$ is density of SC electrons, $n$ is total electron density). In the same time, at $T\gtrsim T_{c1},T_{c2}$ the condensates cannot be considered as independent, hence we cannot find the speed $\upsilon$ using the above consideration. However, we can extrapolate the relation (\ref{3.17}) to all temperatures for dimensional reasons, where the dimensionless coefficient $\chi$ should be considered as an adjustable parameter. So, in general case, it is obvious to assume $\upsilon\sim\upsilon_{F1},\upsilon_{F2}$ so as to $(\hbar\omega_{0})^{2}=\Delta_{01}\Delta_{02}$ (i.e. $\chi=1$).

\begin{figure}[h]
\includegraphics[width=8.5cm]{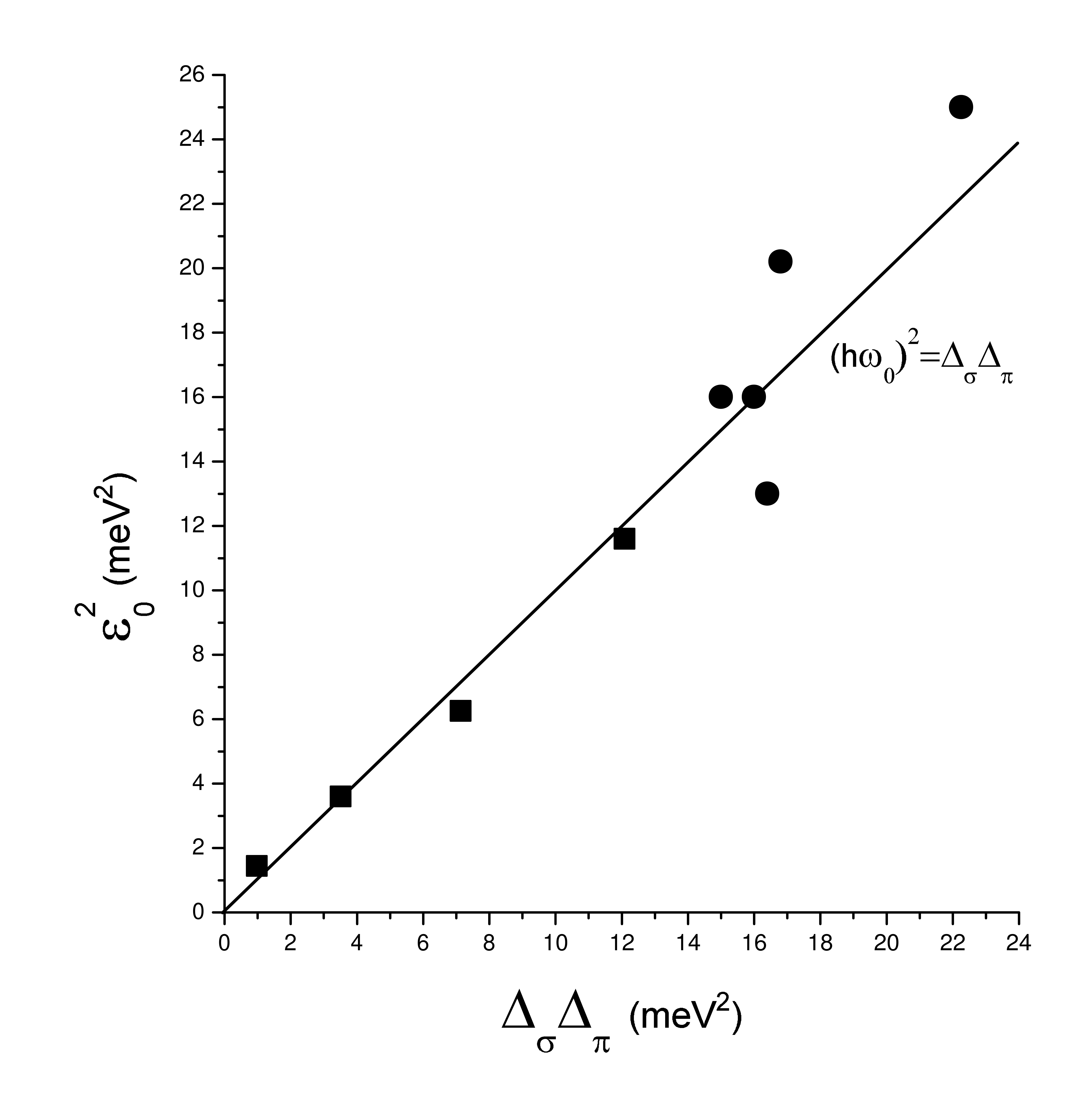}
\caption{Some exitation energy squared $\varepsilon_{0}^{2}$ vs. the product of the gaps $\Delta_{\sigma}\Delta_{\pi}$ at $T=4.2\mathrm{K}$ for $\mathrm{Mg}_{1-x}\mathrm{Al}_{x}\mathrm{B}_{2}$ polycrystalline samples ($6.5\mathrm{K}\leq T_{c}\leq 21.5\mathrm{K}$) - square symbols, and $\mathrm{Mg}\mathrm{B}_{2}$ polycrystalline samples ($28\mathrm{K}\leq T_{c}\leq 40\mathrm{K}$) - circle symbols, the solid line - theoretical energy of the Higgs mode squared vs. the product of the gaps: $(\hbar\omega_{0})^{2}=\Delta_{\sigma}\Delta_{\pi}$ - Eq.(\ref{3.17}) with $\chi=1$.}
\label{Fig2}
\end{figure}

In Ref.\cite{pono2} the measurements of the excitation energy squared $\varepsilon_{0}^{2}$ vs. the product of the gaps $\Delta_{\sigma}\Delta_{\pi}$ have been done for polycrystalline samples $\mathrm{Mg}_{1-x}\mathrm{Al}_{x}\mathrm{B}_{2}$ and $\mathrm{Mg}\mathrm{B}_{2}$ with different $T_{c}$ at the same temperature. The results of the measurements are shown in Fig.(\ref{Fig2}). We can see, that the dependence $\varepsilon_{0}^{2}=\Delta_{\sigma}\Delta_{\pi}$ takes place for both substances, that corresponds to our theoretical result $(\hbar\omega_{0})^{2}=\Delta_{\sigma}\Delta_{\pi}$ for energy of the Higgs mode - Eq.(\ref{3.17}) with $\chi=1$. If we interpret this result as the Leggett mode, then in a formula from Ref.\cite{shar}: $\epsilon_{0}^{2}=4\Delta_{\sigma}\Delta_{\pi}\left[(\lambda_{12}+\lambda_{21})/(\lambda_{11}\lambda_{22}-\lambda_{12}\lambda_{21})\right]$, where $\lambda_{ij}$ is dimensionless interband and intraband coupling constants, we must suppose $(\lambda_{12}+\lambda_{21})/(\lambda_{11}\lambda_{22}-\lambda_{12}\lambda_{21})=1/4$ for different materials with different $T_{c}$, hence with various $\lambda_{ij}$, that is very unlikely.

\section{Effect of Higgs oscillations on Josephson current}\label{Josephson}

In Ref.\cite{pono1,pono2} a resonant enhancement of the DC current through a Josephson junction (JJ) at a bias voltage $V_{\mathrm{res}}$ has been observed when the Josephson frequency $\omega_{J}=\frac{2e}{\hbar}V$ or its harmonics $(m\omega_{J})$ match the energy of some internal oscillation mode of two-band superconductor $\omega_{0}$ or its harmonics $n\omega_{0}$:
\begin{equation}\label{4.1}
  V_{\mathrm{res}}=\frac{n}{m}\frac{\hbar\omega_{0}}{2e},
\end{equation}
where $n$ and $m$ are integer numbers. For $\mathrm{MgB}_{2}$ the value of the energy is observed as $\hbar\omega_{0}\simeq 4\mathrm{meV}$ and $n/m=3/2,1/1,1/2,1/3$. This phenomenon has been observed as the dips in the $\mathrm{d}I/\mathrm{d}V$ characteristics at the voltages $V_{\mathrm{res}}$. In general, peculiarities of the same type could also appear due to interaction of the AC Josephson current with phonons \cite{maks,schl,pono3} or electromagnetic waves \cite{kulik}. Nevertheless, the authors in Ref.\cite{pono1,pono2} believe, that the peculiarities observed in their investigation are related namely to the Leggett collective excitations, based on a work \cite{agter}, where it has been shown, that the Leggett collective mode resonantly couples to the AC Josephson current of a junction between a two-band superconductor and a single-band superconductor, so that, when the voltage matches the energy of the Leggett mode, the resonant enhancement of the DC current takes place. In their experimental samples, there are no optical phonons with energy as low as $4\mathrm{meV}$ in $\mathrm{MgB}_{2}$. The effective interaction between Josephson current and low-energy acoustic phonons as well as electromagnetic waves can only exist in the presence of a resonator system inside the junction. Then, the observed subgap structure could appear at voltages matching the energies of resonator eigenmodes. It is very unlikely, that all their break-junctions demonstrating the discussed subgap structure possess identical resonator systems.

Thus, the resonant enhancement of the DC current takes place if an additional coherence is introduced into the JJ with AC Josephson current (resonance with phonons, el.-mag. waves, the Leggett mode etc.). In Sect.\ref{Goldstone and Higgs} we could see, that, as in single-band superconductors, in two-band superconductors one Goldstone mode (unobservable due to absorbtion into el.-mag. field) and only one Higgs mode exists, at the same time, the Leggett mode is prohibited. Let us consider a S–I–S junction, where the right and left banks are the same isotropic s-wave two-band superconductors with the energy gaps $\Delta_{1,2}$. We assume the capacity $C$ of the JJ is small, so that a McCumber parameter is $\beta=(2e/\hbar)I_{m}CR^{2}\ll 1$, that is displacement current is negligible $I_{D}=C\frac{\mathrm{d}V}{\mathrm{d}t}\ll I_{m}$, and I-V characteristic is $V=R\sqrt{I^{2}-I_{m}^{2}}$.

Calculation of the Josephson current between two-band superconductors with the strong impurity intraband scattering rates (the dirty limit described with Usadel equations) and the weak interband scattering in Ref.\cite{yerin,yerin2} leads to:
\begin{eqnarray}\label{4.2}
 I=\frac{\pi|\Delta_{1}|}{2eR_{N1}}\cos\frac{\theta}{2}\mathrm{Arctanh}\sin{\frac{\theta}{2}}+
 \frac{\pi|\Delta_{2}|}{2eR_{N2}}\cos\frac{\theta}{2}\mathrm{Arctanh}\sin{\frac{\theta}{2}}.
\end{eqnarray}
Here, $R_{N1}$ and $R_{N2}$ are the contributions of the normal resistance for each band, $\theta\equiv\theta_{1}^{R}-\theta_{1}^{L}=\theta_{2}^{R}-\theta_{2}^{L}$ is the phase difference between the banks R and L. The interband scattering is neglected in the present approximation, so that the anharmonicity in Eq.(\ref{4.2}) is caused with the strong impurity intraband scattering \cite{yerin,yerin2,asker6}; accordingly, in the pure limit we will have the usual harmonic expression $J=\frac{\pi|\Delta_{1}|}{2eR_{N1}}\sin\theta+\frac{\pi|\Delta_{2}|}{2eR_{N2}}\sin\theta$. Using Eqs.(\ref{3.16}), Eq.(\ref{4.2}) can be rewritten as follows:
\begin{eqnarray}\label{4.3}
 I=\frac{2\pi^{2}T_{c1}|\Psi_{1}|}{eR_{N1}\sqrt{14\zeta(3)n_{1}}}\cos\frac{\theta}{2}\mathrm{Arctanh}\sin{\frac{\theta}{2}}+
 \frac{2\pi^{2}T_{c2}|\Psi_{2}|}{eR_{N2}\sqrt{14\zeta(3)n_{2}}}\cos\frac{\theta}{2}\mathrm{Arctanh}\sin{\frac{\theta}{2}}.
\end{eqnarray}
Since $\sin{\frac{\theta}{2}}\leq 1$ (for maximal current we can suppose $\theta\approx\pi/2$) we can expand $\mathrm{Arctanh}\sin{\frac{\theta}{2}}$ in the series as:
\begin{eqnarray}\label{4.4}
 \cos\frac{\theta}{2}\mathrm{Arctanh}\sin{\frac{\theta}{2}}&\approx&
 \cos\frac{\theta}{2}\left[\sin{\frac{\theta}{2}}+\frac{1}{2}\sin^{3}{\frac{\theta}{2}}
 +\frac{1}{5}\sin^{5}{\frac{\theta}{2}}+\ldots\right]\nonumber\\
 &=&\frac{1}{2}\sin{\theta}-\frac{1}{16}\sin{2\theta}+\frac{1}{8}\sin{\theta}+
 \frac{1}{5\cdot 32}\sin{3\theta}-\frac{1}{40}\sin{2\theta}+\frac{1}{32}\sin{\theta}+\ldots
\end{eqnarray}
Then,
\begin{equation}\label{4.5}
  I\approx I_{m1}\sin{\theta}-I_{m2}\sin{2\theta}+I_{m3}\sin{3\theta}+\ldots,
\end{equation}
where
\begin{equation}\label{4.6}
  I_{m1}=\frac{4\pi^{2}T_{c1}|\Psi_{1}|}{3eR_{N1}\sqrt{14\zeta(3)n_{1}}}
  +\frac{4\pi^{2}T_{c2}|\Psi_{2}|}{3eR_{N2}\sqrt{14\zeta(3)n_{2}}}\equiv I_{m1}^{(1)}+I_{m1}^{(2)},
\end{equation}
at that $I_{m3}\sim 0.1I_{m2}\sim 0.01I_{m1}$.

\begin{figure}[h]
\includegraphics[width=13.0cm]{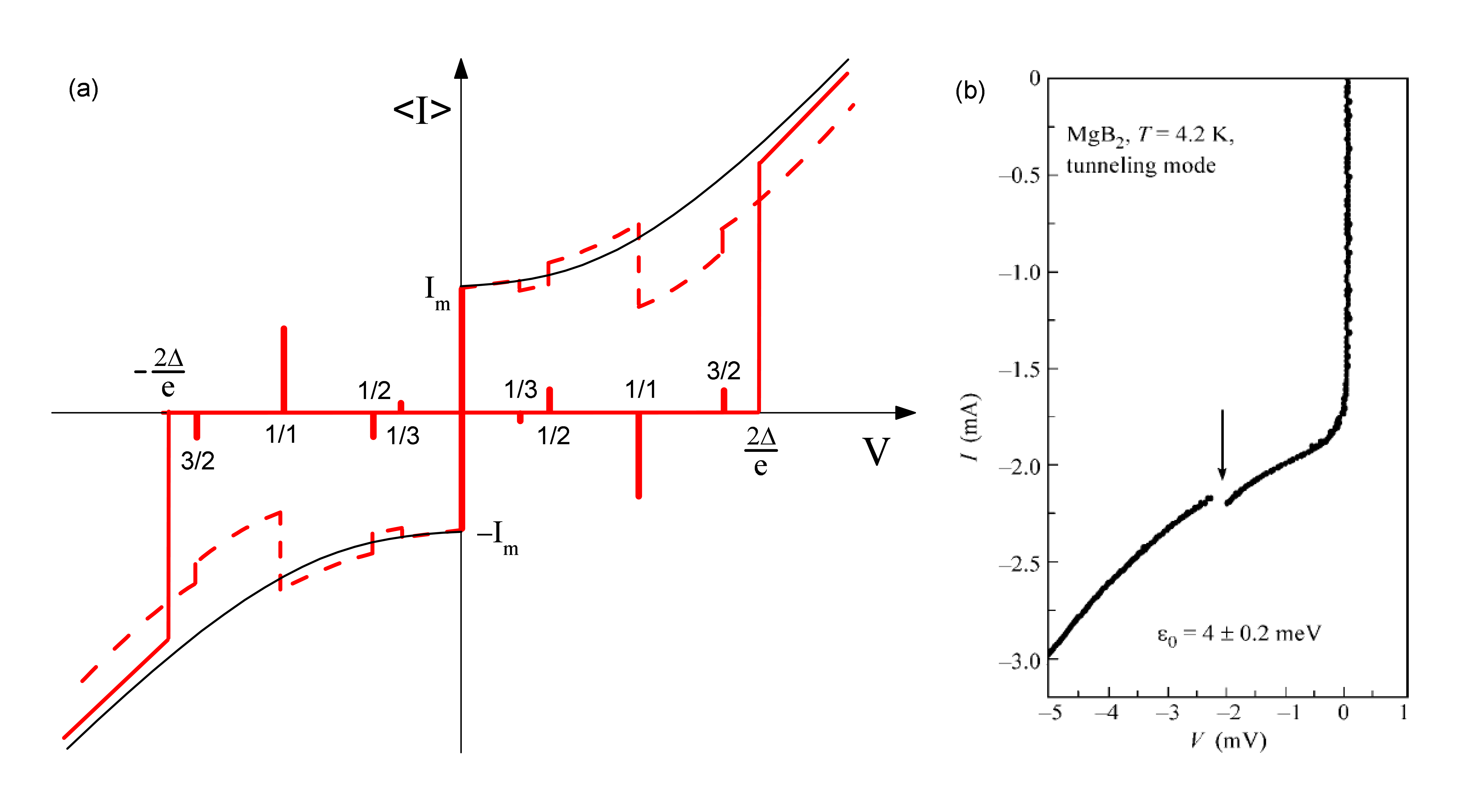}
\caption{(a) - The schematic I-V characteristics of the JJ between two-band superconductors at given voltage (solid line) and at given current (dotted line) as a result of the resonant coupling of AC Josephson current with Higgs oscillations. The resonant subharmonics $\frac{n}{m}=\frac{3}{2},\frac{1}{2},\frac{1}{3}$ from Eq.(\ref{4.17}) are shown for illustration purposes. The thin black solid line is the I-V characteristic $V=R\sqrt{I^{2}-I_{m}^{2}}$ at given current without the coupling. (b) - The fragment of the I-V characteristic of a break junction in a $\mathrm{MgB}_{2}$ sample at $T=4.2\mathrm{K}$ taken from Ref.\cite{pono2}. The structure marked by an arrow is caused by the coupling of AC Josephson current to some oscillation mode with energy $\varepsilon_{0}=4\mathrm{meV}$.}
\label{Fig3}
\end{figure}

Let us consider the problem with the given voltage $V$. Then, in order to ensure the gauge invariance, the phase $\theta$ must be changed as: $\theta\rightarrow\theta-\frac{2e}{\hbar}Vt\equiv\theta-\omega t$, where $\omega\equiv\frac{2e}{\hbar}V$. Then,
\begin{equation}\label{4.7}
  I=-I_{m1}\sin{(\omega t-\theta)}+I_{m2}\sin{(2\omega t-2\theta)}-I_{m3}\sin{(3\omega t-3\theta)}+\ldots
\end{equation}
As in Sect.\ref{Goldstone and Higgs}, we consider the small variations of modulus of the OP from its equilibrium value: $|\Psi_{1,2}|=\Psi_{01,02}+\phi_{1,2}$, where $|\phi_{1,2}|\ll\Psi_{01,02}$. Thus,
\begin{eqnarray}\label{4.8}
  I&=&-I_{m1}\sin{(\omega t-\theta)}-\frac{I_{m1}^{(1)}}{\Psi_{01}}\phi_{1}\sin{(\omega t-\theta)}
  -\frac{I_{m1}^{(2)}}{\Psi_{02}}\phi_{2}\sin{(\omega t-\theta)}\nonumber\\
  &+&I_{m2}\sin{(2\omega t-2\theta)}+\frac{I_{m2}^{(1)}}{\Psi_{01}}\phi_{1}\sin{(2\omega t
  -2\theta)}+\frac{I_{m2}^{(2)}}{\Psi_{02}}\phi_{2}\sin{(2\omega t-2\theta)}\nonumber\\
  &-&I_{m3}\sin{(3\omega t-3\theta)}-\frac{I_{m3}^{(1)}}{\Psi_{01}}\phi_{1}\sin{(3\omega t-3\theta)}
  -\frac{I_{m3}^{(2)}}{\Psi_{02}}\phi_{2}\sin{(3\omega t-3\theta)}+\ldots
\end{eqnarray}
The AC current with frequencies $\omega,2\omega,3\omega,\ldots$ stipulates oscillations of SC density $n_{\mathrm{s}}$ in each band, and besides, according to the result of Sect.\ref{Goldstone and Higgs}, the oscillations in each band occur in phase (in the linear approximation):
\begin{eqnarray}
  e\frac{\partial n_{\mathrm{s1}}}{\partial t}&=&2e\frac{\partial|\Psi_{1}|^{2}}{\partial t}=4e\Psi_{01}\frac{\partial\phi_{1}}{\partial t}=
  -I_{m1}^{(1)}\sin{(\omega t-\theta)}+I_{m2}^{(1)}\sin{(2\omega t-2\theta)}-I_{m3}^{(1)}\sin{(3\omega t-3\theta)}+\ldots\label{4.9a}\\
  e\frac{\partial n_{\mathrm{s2}}}{\partial t}&=&2e\frac{\partial|\Psi_{2}|^{2}}{\partial t}=4e\Psi_{02}\frac{\partial\phi_{2}}{\partial t}=
  -I_{m1}^{(2)}\sin{(\omega t-\theta)}+I_{m2}^{(2)}\sin{(2\omega t-2\theta)}-I_{m3}^{(2)}\sin{(3\omega t-3\theta)}+\ldots\label{4.9b}
\end{eqnarray}
Then we can write the equations:
\begin{eqnarray}
 \frac{\partial^{2}\phi_{1}}{\partial t^{2}}&=&-\frac{I_{m1}^{(1)}\omega}{4e\Psi_{01}}\cos{(\omega t-\theta)}
 +\frac{I_{m2}^{(1)}\omega}{2e\Psi_{01}}\cos{(2\omega t-2\theta)}-\frac{3I_{m3}^{(1)}\omega}{4e\Psi_{01}}\cos{(3\omega t-3\theta)}
 +\ldots\label{4.10a}\\
 \frac{\partial^{2}\phi_{2}}{\partial t^{2}}&=&-\frac{I_{m1}^{(2)}\omega}{4e\Psi_{02}}\cos{(\omega t-\theta)}
 +\frac{I_{m2}^{(2)}\omega}{2e\Psi_{02}}\cos{(2\omega t-2\theta)}-\frac{3I_{m3}^{(2)}\omega}{4e\Psi_{02}}\cos{(3\omega t-3\theta)}
 +\ldots\label{4.10b}
\end{eqnarray}
Thus, if there were no eigen oscillations (Higgs oscillations with the frequency $\omega_{0}$ - Eqs.(\ref{3.15},\ref{3.17})), then the densities of SC electrons $n_{\mathrm{s1}}$ and $n_{\mathrm{s2}}$ oscillate with the frequencies of AC Josephson current $\omega,2\omega,3\omega,\ldots$. On the other hand, the densities $n_{\mathrm{s1}}$ and $n_{\mathrm{s2}}$ can oscillate with the eigen frequency $\omega_{0}$ at some attenuation constant $\gamma\ll\omega_{0}$, and the AC Josephson current plays a role of a driving force. In this system a resonance occurs, if the frequency of AC current $\omega=2eV/\hbar$ coincides with the eigen frequency $\omega_{0}$, then Higgs oscillations in the banks can be excited. In other words, a Cooper pair can tunnel through the JJ, exciting a quantum of Higgs oscillations. Moreover, the tunneling with excitation of $n$ quanta with the total energy $n\hbar\omega_{0}$ by a Cooper pair due to passing through the energy difference $2eV$ can occur, that is equivalent to the resonant excitation of an oscillator with an eigen frequency $n\omega_{0}$ by the external driving force with the frequency $\omega=\frac{2e}{\hbar}V$. Hence, equations for such oscillations take the following form:
\begin{eqnarray}
 \frac{\partial^{2}\phi_{1}}{\partial t^{2}}+2\gamma\frac{\partial\phi_{1}}{\partial t}+[n\omega_{0}]^{2}\phi_{1}
&=&-\frac{I_{m1}^{(1)}\omega}{4e\Psi_{01}}\cos{(\omega t-\theta)}
 +\frac{I_{m2}^{(1)}\omega}{2e\Psi_{01}}\cos{(2\omega t-2\theta)}-\frac{3I_{m3}^{(1)}\omega}{4e\Psi_{01}}\cos{(3\omega t-3\theta)}
 +\ldots\label{4.11a}\\
 \frac{\partial^{2}\phi_{2}}{\partial t^{2}}+2\gamma\frac{\partial\phi_{2}}{\partial t}+[n\omega_{0}]^{2}\phi_{2}
&=&-\frac{I_{m1}^{(2)}\omega}{4e\Psi_{02}}\cos{(\omega t-\theta)}
 +\frac{I_{m2}^{(2)}\omega}{2e\Psi_{02}}\cos{(2\omega t-2\theta)}-\frac{3I_{m3}^{(2)}\omega}{4e\Psi_{02}}\cos{(3\omega t-3\theta)}
 +\ldots\label{4.11b}
\end{eqnarray}
The particular solutions of nonhomogeneous differential equations are (the general solutions are attenuating, therefore they can be omitted):
\begin{eqnarray}\label{4.12}
&&\phi_{1,2}\nonumber\\
&&=\frac{I_{m1}^{(1,2)}\omega
\cos\left(\omega t-\theta-\varphi_{1}\right)}{4e\Psi_{01,02}\sqrt{\left(\omega^{2}-[n\omega_{0}]^{2}\right)^{2}+4\gamma^{2}\omega^{2}}}
  +\frac{I_{m2}^{(1,2)}\omega
  \cos\left(2\omega t-2\theta-\varphi_{2}\right)}{2e\Psi_{01,02}\sqrt{\left(4\omega^{2}-[n\omega_{0}]^{2}\right)^{2}+16\gamma^{2}\omega^{2}}}
  +\frac{3I_{m3}^{(1,2)}\omega
  \cos\left(3\omega t-3\theta-\varphi_{3}\right)}{4e\Psi_{01,02}\sqrt{\left(9\omega^{2}-[n\omega_{0}]^{2}\right)^{2}+36\gamma^{2}\omega^{2}}}\nonumber\\
  &&+\ldots,
\end{eqnarray}
where
\begin{equation}\label{4.13}
  \cos\varphi_{1}=\frac{\omega^{2}-[n\omega_{0}]^{2}}{\sqrt{\left(\omega^{2}-[n\omega_{0}]^{2}\right)^{2}+4\gamma^{2}\omega^{2}}},\quad
  \cos\varphi_{2}=\frac{[n\omega_{0}]^{2}-4\omega^{2}}{\sqrt{\left(4\omega^{2}-[n\omega_{0}]^{2}\right)^{2}+16\gamma^{2}\omega^{2}}},\quad
  \cos\varphi_{3}=\frac{9\omega^{2}-[n\omega_{0}]^{2}}{\sqrt{\left(9\omega^{2}-[n\omega_{0}]^{2}\right)^{2}+36\gamma^{2}\omega^{2}}}.
\end{equation}
We can see, that resonance occurs at frequencies $\omega=n\omega_{0},n\omega_{0}/2,n\omega_{0}/3,\ldots$ (if $\gamma\ll\omega_{0}$). In the resonant frequencies we have $\varphi_{1,2,3}=\pi/2$. Then the solution (\ref{4.12}) in the resonant frequencies is
\begin{eqnarray}\label{4.14}
  \phi_{1,2}\left(n\omega_{0}\right) &=& \frac{I_{m1}^{(1,2)}}{8e\Psi_{01,02}\gamma}\sin\left(n\omega_{0} t-\theta\right) \nonumber\\
  \phi_{1,2}\left(\frac{n\omega_{0}}{2}\right) &=& \frac{I_{m2}^{(1,2)}}{8e\Psi_{01,02}\gamma}\sin\left(n\omega_{0} t-2\theta\right) \\
  \phi_{1,2}\left(\frac{n\omega_{0}}{3}\right) &=& \frac{I_{m3}^{(1,2)}}{8e\Psi_{01,02}\gamma}\sin\left(n\omega_{0} t-3\theta\right). \nonumber
\end{eqnarray}
Out of the resonances the amplitudes of $\phi_{1,2}$ are negligible. Substituting the solutions (\ref{4.14}) in the current (\ref{4.8}) we obtain the currents at the corresponding resonant frequencies:
\begin{eqnarray}\label{4.15}
  I(n\omega_{0})&=&-I_{m1}\sin{(n\omega_{0} t-\theta)}+I_{m2}\sin{(2n\omega_{0} t-2\theta)}-I_{m3}\sin{(3n\omega_{0} t-3\theta)}\nonumber\\
  &+&\frac{\left[I_{m1}^{(1)}\right]^{2}}{16e\Psi_{01}^{2}\gamma}\cos{(2n\omega_{0} t-2\theta)}
  +\frac{\left[I_{m1}^{(2)}\right]^{2}}{16e\Psi_{02}^{2}\gamma}\cos{(2n\omega_{0} t-2\theta)}
  -\frac{\left[I_{m1}^{(1)}\right]^{2}}{16e\Psi_{01}^{2}\gamma}-\frac{\left[I_{m1}^{(2)}\right]^{2}}{16e\Psi_{02}^{2}\gamma}+\ldots\nonumber\\
  I\left(\frac{n}{2}\omega_{0}\right)&=&-I_{m1}\sin{\left(\frac{n}{2}\omega_{0} t-\theta\right)}+I_{m2}\sin{\left(n\omega_{0} t-2\theta\right)}-I_{m3}\sin{\left(\frac{3n}{2}\omega_{0} t-3\theta\right)}\nonumber\\
  &-&\frac{\left[I_{m2}^{(1)}\right]^{2}}{16e\Psi_{01}^{2}\gamma}\cos{(2n\omega_{0} t-4\theta)}
  -\frac{\left[I_{m2}^{(2)}\right]^{2}}{16e\Psi_{02}^{2}\gamma}\cos{(2n\omega_{0} t-4\theta)}
  +\frac{\left[I_{m2}^{(1)}\right]^{2}}{16e\Psi_{01}^{2}\gamma}+\frac{\left[I_{m2}^{(2)}\right]^{2}}{16e\Psi_{02}^{2}\gamma}+\ldots\nonumber\\
  I\left(\frac{n}{3}\omega_{0}\right)&=&-I_{m1}\sin{\left(\frac{n}{3}\omega_{0} t-\theta\right)}+I_{m2}\sin{\left(\frac{2n}{3}\omega_{0} t-2\theta\right)}-I_{m3}\sin{\left(n\omega_{0} t-3\theta\right)}\nonumber\\
  &+&\frac{\left[I_{m3}^{(1)}\right]^{2}}{16e\Psi_{01}^{2}\gamma}\cos{(2n\omega_{0} t-6\theta)}
  +\frac{\left[I_{m3}^{(2)}\right]^{2}}{16e\Psi_{02}^{2}\gamma}\cos{(2n\omega_{0} t-6\theta)}
  -\frac{\left[I_{m3}^{(1)}\right]^{2}}{16e\Psi_{01}^{2}\gamma}-\frac{\left[I_{m3}^{(2)}\right]^{2}}{16e\Psi_{02}^{2}\gamma}+\ldots
\end{eqnarray}
Out of the resonances the current has the form:
\begin{equation}\label{4.16}
  I(\omega)=-I_{m1}\sin{(\omega t-\theta)}+I_{m2}\sin{(2\omega t-2\theta)}-I_{m3}\sin{(3\omega t-3\theta)}+\ldots
\end{equation}
We can see, that \emph{at the frequencies $\omega=n\omega_{0},n\omega_{0}/2,n\omega_{0}/3,\ldots$ the DC current occurs, however, unlike DC Josephson current} (\ref{4.2},\ref{4.5})\emph{, it is not function of the phase difference $\theta$}. Thus, at the voltage
\begin{equation}\label{4.17}
V_{\mathrm{res}}=\frac{n}{m}\frac{\hbar\omega_{0}}{2e},
\end{equation}
where $m,n=1,2,3\ldots$, DC current through the JJ appears. Amplitudes of the resonant peaks is proportional to the square of the critical Josephson currents $I_{m1}^{2},I_{m2}^{2},I_{m3}^{2},\ldots$. As we have seen above:  $I_{m3}\sim 0.1I_{m2}\sim 0.01I_{m1}$, hence, DC resonant currents are such, that $I\left(\frac{1}{3}\omega_{0}\right)\sim 10^{-2}I\left(\frac{1}{2}\omega_{0}\right)\sim 10^{-4}I(\omega_{0})$, as we can see from Eq.(\ref{4.15}). Thus, the main resonance at the voltage $V_{\mathrm{res}}=\frac{\hbar\omega_{0}}{2e}$ (that is, when $\omega=\omega_{0}$) is the most pronounced. In I-V characteristic of $\mathrm{Mg}\mathrm{B}_{2}$ break junction (S-I-S type) a "step" has been observed in Ref.\cite{pono2} just at voltage $V_{\mathrm{res}}=\frac{\hbar\omega_{0}}{2e}$ - Fig.(\ref{Fig3}b).  In the same time, the currents in the subharmonics $\omega=\omega_{0}/2,\omega_{0}/3$ are very weak. In the I-V characteristics of a JJ between two-bands superconductors shown in Fig.(\ref{Fig3}a) we should observe the picture, as Shapiro spikes, at given voltage and, as Shapiro steps, at given current.

\begin{figure}[h]
\includegraphics[width=8.0cm]{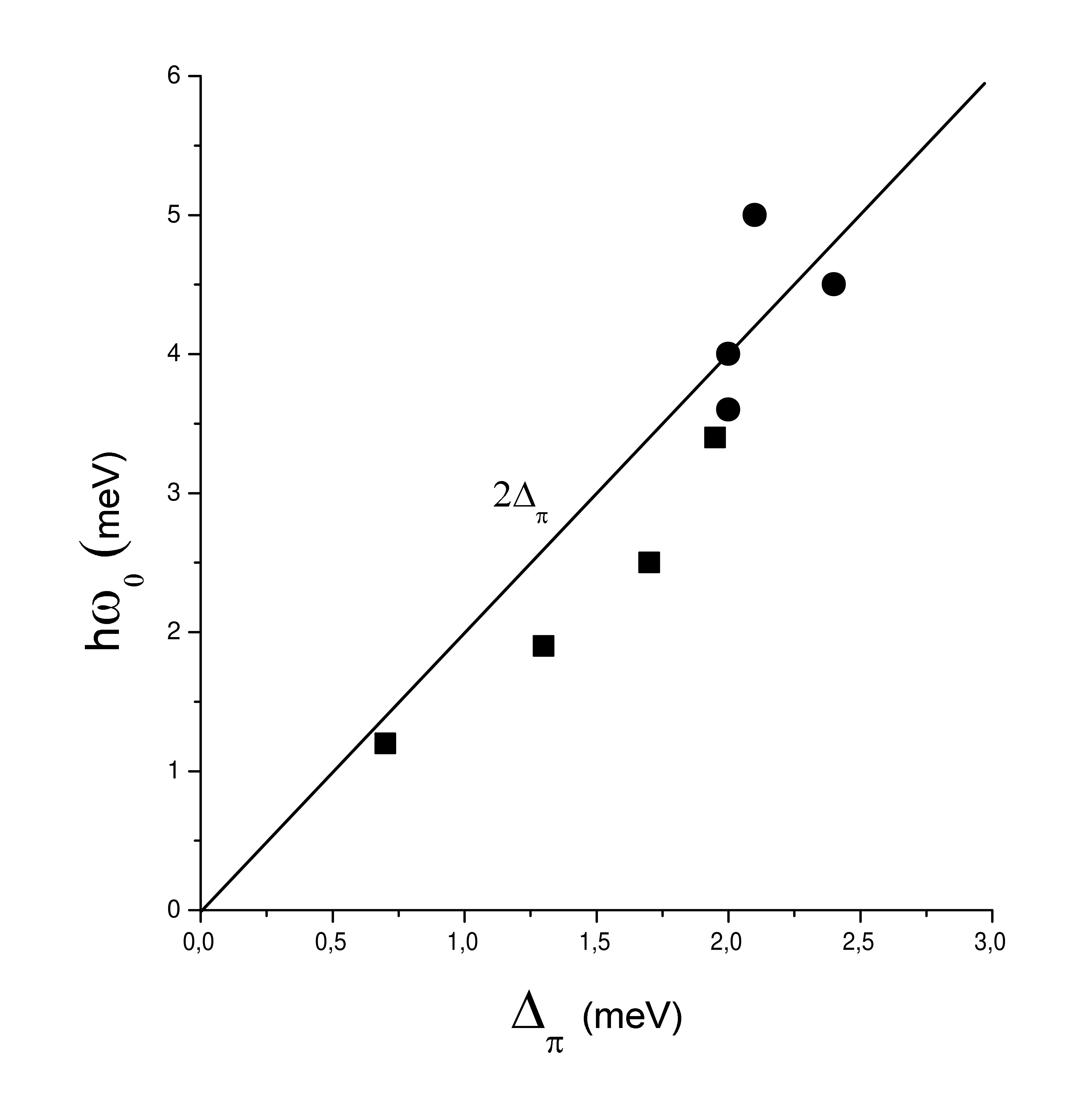}
\caption{The excitation energy $\hbar\omega_{0}$ vs. the minimal gap $\Delta_{\pi}$ at $T=4.2\mathrm{K}$ for $\mathrm{Mg}_{1-x}\mathrm{Al}_{x}\mathrm{B}_{2}$ polycrystalline samples ($6.5\mathrm{K}\leq T_{c}\leq 21.5\mathrm{K}$) - square symbols, and $\mathrm{Mg}\mathrm{B}_{2}$ polycrystalline samples ($28\mathrm{K}\leq T_{c}\leq 40\mathrm{K}$) - circle symbols, the solid line is $2\Delta_{\pi}$, below which the resonant enhancement of the DC current through the JJ can be observable.}
\label{Fig4}
\end{figure}

Thus, the resonant enhancement of the DC current through the JJ is result of the coupling between AC Josephson current and Higgs oscillations. It should be noted, that we obtain the result (\ref{4.15}) neglecting of the interband scattering, hence, the effect would have been observed also in single-band superconductors. However, in the single-band superconductors the minimal energy of the Higgs mode is $\hbar\omega_{0}=2|\Delta|$ \cite{grig2}, that is this mode exists in the free quasiparticle continuum. Hence, Higgs oscillations decay to quasiparticles with energy $\sqrt{|\Delta|^{2}+v_{F}^{2}(p-p_{F})^{2}}$ each. In other words, excitation of the Higgs mode is accompanied by the breaking of Cooper pairs with transfer of their constituents in the free quasiparticle states. Thus, the Higgs mode is unstable in single-band superconductors. Therefore, the resonant enhancement is strongly suppressed. On the contrary, as shown in Sect.\ref{Goldstone and Higgs}, in two-band superconductors $(\hbar\omega_{0})^{2}=|\Delta_{1}||\Delta_{2}|$ takes place. If $\hbar\omega_{0}<2\min(|\Delta_{1}|,|\Delta_{2}|)$, then the Higgs mode is stable, therefore, the corresponding resonances in AC Josephson effect can be observable. The subharmonics at higher frequencies $\omega=\frac{n}{m}\omega_{0}>\omega_{0}$ can be observed if $V_{\mathrm{res}}=\frac{\hbar\omega}{2e}<2\min(|\Delta_{1}|,|\Delta_{2}|)/e$; at $V_{\mathrm{res}}\geq 2\min(|\Delta_{1}|,|\Delta_{2}|)/e$ the Josephson current $I(V)$ quickly goes to zero, and the total current is the tunneling of quasiparticles through the contact \cite{till}. Using experimental data of \cite{pono2} we can verify this criterion, which has a form $\hbar\omega_{0}\lesssim 2\Delta_{\pi}$ in this case. From Fig.(\ref{Fig4}) we can see, that the criterion is approximately satisfied.


\section{Results}\label{results}

In this work we investigate the eigen oscillations of internal degrees of freedom of two-band superconductors using the action (\ref{1.6}) with the corresponding Lorentz-invariant Lagrangian (\ref{1.8}), which are generalization of the GL free energy functional for a two-band isotropic s-wave superconductor on the one hand, and of the ETDGL theory formulated in Ref.\cite{grig2} on the other hand. Our results are as follows:

1) Due to the internal proximity effect, the Goldstone mode splits into two branches: common mode oscillations with an acoustic spectrum - Eq.(\ref{2.7}), and the oscillations of the relative phase $\theta_{1}-\theta_{2}$ between two SC condensates with an energy gap in spectrum determined by the interband coupling $\epsilon$ - Eqs.(\ref{2.10},\ref{2.12}), which can be associated with the Leggett mode. The common mode oscillations are absorbed into the gauge field $A_{\mu}$ due to these oscillations are accompanied by current, as in single-band superconductors \cite{grig2}. At the same time, Leggett oscillations are not accompanied by the current, therefore they "survive". Higgs oscillations splits into two branches also: a massive one, whose energy gap (mass) vanishes at $T_{c}$ - Eq.(\ref{3.10a}), another massive one, whose mass, determined by the interband coupling $\epsilon$, does not vanish at $T_{c}$ - Eqs.(\ref{3.10b},\ref{3.10c}). For the first mode, oscillations of $|\Psi_{1}|$ and $|\Psi_{2}|$ occur in a phase, for the second mode, the oscillations occur in an anti-phase. The mass of the Higgs mode is related to the coherence length $\xi$, hence, we obtain two coherence lengths accordingly to the branches - Eqs.(\ref{3.11a},\ref{3.11b}). The first coherence length diverges at $T=T_{c}$, on the contrary, the second length remains finite at all temperatures. It should be noted, that the effect of the splitting of Goldstone and Higgs modes into two branches each takes place even at the infinitely small coefficient $\epsilon$. Thus, \emph{the effect of interband coupling $\epsilon\neq 0$, even if the coupling is weak $|\epsilon|\ll |a_{1,2}(0)|$, is nonperturbative}.

2) To excite one quanta of Higgs oscillations, as least one Cooper pair must be broken, i.e. the energy of order of $2|\Delta|$ must be spent. Thus, excitations of any Higgs mode at $T=T_{c}$ does not require the energy consumption, since $|\Delta|=0$. In two-band superconductors for the anti-phase Higgs mode - Eqs.(\ref{3.10b},\ref{3.10c}) we have the nonphysical property $q_{\mu}q^{\mu}(T_{c})\neq 0$. If $\eta^{2}=\frac{1}{m_{1}m_{2}},\quad\eta\epsilon<0$, then from Eq.(\ref{3.7}) we can see, that the anti-phase Higgs mode is absent, and the common mode oscillations with the zero energy gap at $T=T_{c}$ - Eq.(\ref{3.11}) remains only. As a consequence, only the coherence length (\ref{3.11a}) remains, that prohibits type 1.5 superconductors. Analogously, from Eq.(\ref{2.6}) we can see, that the Leggett mode is absent, and only the common mode oscillations with the gapless spectrum (\ref{2.7}) remains. Therefore, we must suppose Eq.(\ref{3.14}), that ensures the correct property of the Higgs mode. \emph{Thus, as in single-band superconductors, in two-band superconductors only one Goldstone mode and one Higgs mode exist, and the Leggett mode and the anti-phase Higgs mode are absent}. However, as mentioned above, the Goldstone mode is accompanied by the current, therefore the gauge field $\widetilde{A}_{\mu}$ absorbs the Goldstone boson $\theta$, as in single-band superconductors.

3) The energy gap of the Higgs mode in two-band superconductors - Eq.(\ref{3.11}) can be represent in the form: $\hbar\omega_{0}=\sqrt{|\Delta_{1}||\Delta_{2}|}$, that differs from the mass of the Higgs mode in single-band superconductors $\hbar\omega_{0}=2|\Delta|$. Thus, in single-band superconductors this mode exists in the free quasiparticle continuum, hence, it is unstable. On the contrary, in two-band superconductors it can be $\sqrt{|\Delta_{1}||\Delta_{2}|}<2\min(|\Delta_{1}|,|\Delta_{2}|)$, then the Higgs mode becomes stable. The reason for the difference in energies between the Higgs modes in two-band and single-band SC systems is, that the positive feedback in Eqs.(\ref{3.3}) for equilibrium values of the OP occurs. The "light" speed $\upsilon$ is determined with Eqs.(\ref{3.19},\ref{3.20}), i.e. it is of order of Fermi velocities in the bands: $\upsilon\sim\upsilon_{F1},\upsilon_{F2}$. The result for spectrum of Higgs oscillations in two-band superconductors has been compared with measurements of the excitation energy $\varepsilon_{0}$ for polycrystalline samples $\mathrm{Mg}_{1-x}\mathrm{Al}_{x}\mathrm{B}_{2}$ and $\mathrm{Mg}\mathrm{B}_{2}$ with different $T_{c}$ at the same temperature \cite{pono2} - Fig.(\ref{Fig2}). We can see, that the dependence $\varepsilon_{0}^{2}=\Delta_{\sigma}\Delta_{\pi}$ takes place for both substances, that corresponds to our theoretical result for the Higgs mode. At the same time, these experimental results are difficult to interpret as manifestation of the Leggett mode, based on the formula of Ref.\cite{shar}.

4) It is demonstrated, that the resonant enhancement of the DC current through a JJ at the resonant bias voltage $V_{\mathrm{res}}$, when the Josephson frequency or its harmonics match the frequency of some internal oscillation mode or its harmonics in two-band superconductors (banks) - Eq.(\ref{4.1}), which was observed in the experiments \cite{pono1,pono2}, is a result of coupling between AC Josephson current and Higgs oscillations - Eqs.(\ref{4.15},\ref{4.17}). Amplitudes of the resonant peaks are proportional to the square of the critical Josephson currents. The main resonance at the voltage $V_{\mathrm{res}}=\frac{\hbar\omega_{0}}{2e}$ is the most pronounced. The currents in subharmonics are very weak. These subharmonics are caused with the anharmonicity in the current-phase relation (\ref{4.2}) and with excitation of several quanta of the Higs mode simultaneously by a Cooper pair due to the passing through the energy difference $2eV$.  In the I-V characteristics of the JJ between two-bands superconductors we should observe picture, as Shapiro spikes, at given voltage and, as Shapiro steps, at given current - Fig.(\ref{Fig3}). This phenomenon would have been observed in single-band superconductors also. However, as mentioned before, the Higgs mode is unstable in single-band superconductors. Therefore, the resonant enhancement is strongly suppressed. On the contrary, in two-band superconductors, if $\hbar\omega_{0}<2\min(|\Delta_{1}|,|\Delta_{2}|)$ occurs, then the Higgs mode is stable, therefore, the corresponding resonances in AC Josephson effect can be observable. Thus, \emph{explanation of this effect does not need Leggett oscillations, hence the effect cannot be considered as experimental confirmation of these oscillations}.

\newpage
\acknowledgments


This research was supported by grant of National Research Foundation of Ukraine "Models of nonequilibrium processes in colloidal systems" 2020.02/0220 and by theme grant of Department of physics and astronomy of NAS of Ukraine: "Noise-induced dynamics and correlations in nonequilibrium systems" 0120U101347


\end{document}